\documentclass[letterpaper,journal]{IEEEtran}
\usepackage{amsmath}
\usepackage{amsthm}
\allowdisplaybreaks
\usepackage{amssymb}
\usepackage{autobreak}
\usepackage{array}
\usepackage{algorithmic}
\usepackage{algorithm}
\usepackage{bm}
\usepackage{booktabs}
\usepackage{cite}
\usepackage{enumitem}
\setlist[itemize]{leftmargin=*}
\setlist[enumerate]{leftmargin=*}
\usepackage{graphicx}
\usepackage{epstopdf}
\usepackage{makecell}
\usepackage{multirow} % Add this line
\usepackage{mathtools}
\usepackage{multicol}
\usepackage{paracol}
\usepackage{placeins}
\usepackage{stfloats}
\usepackage{tabularx}
\usepackage{colortbl}
\usepackage[skins]{tcolorbox}
\usepackage{textcomp}
\usepackage{url}
\usepackage{verbatim}
\usepackage{xcolor}
\usepackage{tikz}
\usepackage{pifont} % Add this line for symbols like \ding{51} and \ding{55}
\usepackage{fancyhdr}
\newtheorem{theorem}{Theorem}
\newtheorem{definition}{Definition}
\newtheorem{lemma}{Lemma}

\begin{document}

\title{SilentLedger: Privacy-Preserving Auditing for Blockchains with Complete Non-Interactivity}

\author{Zihan Liu, Xiaohu Wang, Chao Lin*, Minghui Xu*, Debiao He, and Xinyi Huang
        % <-this % stops a space
\thanks{This work was supported in part by the National Key Research and Development Program of China under Grant 2025YFB3110200, the National Natural Science Foundation of China under Grants 62425205, 62572216, U25A20425, 62302266, 62232010, U23A20302, and U24A20244, and the Taishan Scholars Program under Grant tsqn202507069. \textit{(Corresponding authors: Chao Lin; Minghui Xu.)}}
\thanks{Zihan Liu and Minghui Xu are with the School of Artificial Intelligence/School of Computer Science and Technology, Shandong University, Jinan 250100/Qingdao 266237, China. (E-mail: liuzihan990911@outlook.com, mhxu@sdu.edu.cn)}% <-this % stops a space
\thanks{Xiaohu Wang is with the School of Cyber Science and Technology, Beihang University, Beijing 100191, China. (E-mail: by2439130@buaa.edu.cn)}
\thanks{Chao Lin and Xinyi Huang are with the College of Computer Science and Technology/College of Software, Nanjing University of Aeronautics and Astronautics, Nanjing 211106, China. (E-mail: chaolin@nuaa.edu.cn, huangxinyi@nuaa.edu.cn)}
\thanks{Debiao He is with the School of Cyber Science and Engineering, Wuhan University, Wuhan 430072, China. (E-mail: hedebiao@whu.edu.cn)}
}

\markboth{IEEE Transactions on Information Forensics and Security}%
{Liu \MakeLowercase{\textit{et al.}}: SilentLedger: Privacy-Preserving Auditing for Blockchains with Complete Non-Interactivity}

\maketitle

\begin{abstract}
Privacy-preserving blockchain systems are essential for protecting transaction data, yet they must also provide auditability that enables authorized auditors to recover participant identities and transaction amounts when warranted. Existing auditable designs still rely on off-chain interaction between auditors and users during transaction generation or auditing. These runtime dependencies are hard to sustain in open blockchain settings. Moreover, many auditable solutions require auditors to serve as validator nodes, which broadens the exposure of sensitive transaction data and makes the system vulnerable to performance bottlenecks or failures of auditors.

To support off-line transactions and independent auditing, we propose SilentLedger, a privacy-preserving transaction system with auditing and complete non-interactivity. To enable public verification without revealing account identities, we introduce the renewable anonymous certificate scheme with formal syntax and security notions. SilentLedger further combines this certificate mechanism with encryption scheme and signatures of knowledge, enabling payers to generate transactions without payee or auditor participation while allowing authorized auditors to recover audit data solely from on-chain data. We formally prove security properties including authenticity, anonymity, confidentiality, and soundness, and evaluate performance under a standard 2-2 transaction model. Our implementation and benchmarks demonstrate that SilentLedger achieves practical performance compared with recent solutions.
\end{abstract}


\section{Introduction}
\IEEEPARstart{B}{lockchain-based} decentralized transactions have fundamentally transformed financial operations by enabling trustless peer-to-peer payments without intermediaries~\cite{The_truth_about_blockchain}.	However, the inherent transparency of early blockchain systems such as Bitcoin has created privacy vulnerabilities that hinder adoption in financial applications that are sensitive to privacy~\cite{DCAP, Ppchain}. Following Bulletproofs~\cite{Bulletproofs}, the privacy of a blockchain transaction consists of two properties: anonymity, which hides participant identities from public observers~\cite{CryptoNote2}, and confidentiality, which hides transaction amounts. While schemes such as Zerocash~\cite{Zerocash}, RingCT 2.0~\cite{Ringct_2.0}, and Mimblewimble~\cite{Aggregate_cash_systems} provide strong guarantees for both properties, they require interaction between the payer and payee. This is because these designs use cryptographic commitments with embedded randomness, later used as spending keys, that must be shared with the payee to enable future spending.

To eliminate this interactivity, Quisquis~\cite{Quisquis} encrypts the transaction amount under the payee's public key, allowing the payer to prove control of the corresponding secret key without requiring the payee to be online. Similarly, Non-interactive Mimblewimble~\cite{Non-interactive_mimblewimble} builds on Mimblewimble's commitment~\cite{pedersen} structure and integrates a Diffie--Hellman key exchange, allowing the payee to reconstruct the randomness in the commitment independently. In this approach, the payer derives a shared value using an ephemeral key and the payee's public key. The payee can later recover this value using their secret key, enabling them to spend the received coins without any prior coordination.

Despite their privacy advantages, these schemes pose challenges for compliance auditing because they can be abused for illicit activities such as money laundering and tax evasion~\cite{Blockchain_Regulatory_Certainty_Act}. Ensuring auditability, namely the capability for authorized entities to examine transactions for legal compliance, is essential in practical financial systems. To reconcile privacy with compliance requirements, recent work~\cite{Traceable_monero, Platypus, Aca, Peredi, zkCross, Privacy-Preserving_Cryptocurrency_With_Threshold_Authentication_and_Regulation} proposes auditable privacy-preserving transaction systems that selectively disclose participants' registered identities or account information and transaction amounts to authorized auditors while keeping them hidden from the public. However, most existing solutions still involve some interaction between users and auditors. For example, Traceable Monero~\cite{Traceable_monero}, Platypus~\cite{Platypus}, and ACA~\cite{Aca} encrypt user identity information under the auditor's public key, enabling independent identity recovery by the auditor. In contrast, for hidden amounts, Traceable Monero and Platypus rely on homomorphic commitments, whereas ACA employs homomorphic encryption under user-specific keys, both of which require user cooperation to disclose amounts.

Among the schemes that attempt to eliminate interaction during auditing, PGC~\cite{PGC} supports non-interactive auditing of hidden amounts under a privileged supervision model. However, PGC adopts a pseudonymous account model and does not hide participant identities, so it does not satisfy the full privacy-preserving goal. Under a stronger privacy model, PEReDi~\cite{Peredi} encrypts both identity information and transaction amounts under the auditor's public key, supporting homomorphic operations and allowing the auditor to decrypt ciphertexts directly. Although this removes the need for users to participate in disclosure, it introduces a runtime dependency on the auditor during transaction execution. Unlike Traceable Monero and ACA where users can independently verify the encrypted values, PEReDi does not provide this capability for encrypted amounts. Instead, users must rely on the auditor to attest to or sign transaction status updates. Consequently, PEReDi eliminates user interaction during auditing but shifts interaction back to transaction execution, thereby weakening transaction independence.

In this paper, we aim to \textit{achieve complete non-interactivity in privacy-preserving auditing}: users can transact without interaction, and authorized auditors can recover audit information from on-chain data without user cooperation.
This property is crucial for practical deployment. Requiring the payee to be online is often impractical in point-of-sale payments, cross-border remittances, or IoT-based microtransactions. Offline payee support greatly improves usability and broadens deployment in asynchronous or resource-constrained settings. Equally important is offline auditability, enabling auditors to access necessary transaction data independently of user availability. By decoupling interactivity from both transaction execution and auditing, complete non-interactivity enhances usability, reliability, and compliance, which are key pillars for practical adoption.

\begin{table*}[!b]
	\caption{Interactivity and privacy comparison of related auditable payment systems.}
	\label{tab:comparison}
	\centering
	\small
	\setlength{\tabcolsep}{5pt}
	\renewcommand{\arraystretch}{1.15}
	\rowcolors{4}{gray!15}{white}
	\begin{tabular}{lcccccc}
		\toprule
		\multirow{2}{*}{\textbf{Protocol}} &
		\multicolumn{2}{c}{Privacy Mechanisms\textsuperscript{1}} &
		\multicolumn{2}{c}{Transaction\textsuperscript{2}} &
		\multicolumn{2}{c}{Audit\textsuperscript{2}} \\
		\cmidrule(lr){2-3}
		\cmidrule(lr){4-5}
		\cmidrule(lr){6-7}
		& Anonymity & Confidentiality &
		Balance update &
		Transaction posting &
		Identity &
		Amount \\
		\hline
		PGC~\cite{PGC} & $\times$ & Enc & $\bullet$ & $\bullet$ & $\times$ & $\bullet$ \\
		Traceable Monero~\cite{Traceable_monero} & Enc & Com & $\circ$ & $\bullet$ & $\bullet$ & $\times$ \\
		Platypus~\cite{Platypus} & Enc & Com & $\circ$ & $\circ$ & $\bullet$ & $\circ$ \\
		ACA~\cite{Aca} & Enc & Enc & $\bullet$ & $\bullet$ & $\bullet$ & $\circ$ \\
		PEReDi~\cite{Peredi} & Enc & Enc & $\bullet$ & $\circ$ & $\bullet$ & $\bullet$ \\
		\hline
		\textbf{SilentLedger} & Enc & Enc & $\bullet$ & $\bullet$ & $\bullet$ & $\bullet$ \\
		\bottomrule
	\end{tabular}
	{
		\footnotesize
		\begin{itemize}
			\item[1] Enc: Encryption, Com: Commitment.
			\item[2] ($\bullet$) indicates that the corresponding transaction generation or auditing operation is supported without interaction, ($\circ$) indicates that it is supported but requires interaction, and ($\times$) indicates that the corresponding privacy or audit capability is not supported.
		\end{itemize}
	}
\end{table*}

\subsection{Technical Challenges}

Achieving complete non-interactivity requires payees to spend anonymous outputs with hidden amounts without coordinating with counterparties or auditors, while auditors must recover identities and amounts only from on-chain data and audit keys. Direct solutions are inefficient.
A natural approach is to use broadcast encryption~\cite{New_constructions_on_broadcast_encryption_key_pre-distribution_schemes, Combinatorial_Bounds_for_Broadcast_Encryption} or encryption for multiple receivers~\cite{An_efficient_and_secure_multimessage_and_multireceiver_signcryption_scheme, Efficient_anonymous_multireceiver_certificateless_encryption} to encode both anonymous addresses and hidden amounts. However, these primitives incur substantial computational and communication overhead, rendering them impractical for deployments in practice.

An alternative that attaches separate ciphertexts for the payee and auditor avoids complex cryptography for multiple recipients, but it doubles the computation and communication overhead and is difficult to integrate into existing blockchain ecosystems. In particular, public-key encryption over account addresses often requires large plaintext support (e.g., a point on the secp256k1 curve spans 512 bits~\footnote{https://en.bitcoin.it/wiki/Secp256k1}). Traditional schemes like RSA~\cite{RSA} or Paillier~\cite{Paillier} support such sizes but are incompatible with blockchain environments that predominantly use elliptic curve cryptography. Meanwhile, elliptic curve-friendly schemes such as exponential ElGamal~\cite{ElGamal} and exponential linear encryption~\cite{Short_Group_Signatures} only support small plaintext spaces, rendering them incompatible for this use case.

Another key challenge is ensuring that anonymous addresses are verifiably derived from registered account addresses, so adversaries cannot forge identities or generate sham transactions to evade compliance audits. Recent systems such as Platypus~\cite{Platypus} and PEReDi~\cite{Peredi} require auditors to serve as recording nodes to preserve this linkage. However, this design creates a single point of failure for privacy: compromise of any recording node can expose the linkage. Moreover, scaling the number of recording nodes enlarges the attack surface and further elevates the risk of privacy breaches.

\subsection{Our Contributions}

We propose SilentLedger, which is, to our knowledge, the first privacy-preserving auditing system that achieves complete non-interactivity in both transaction execution and auditing. Table~\ref{tab:comparison} compares SilentLedger with state-of-the-art schemes. Our core contributions are summarized as follows:
\begin{enumerate}
	\item \textbf{Complete Non-interactivity.} We present SilentLedger, the first privacy-preserving auditing transaction system to eliminate interaction in both transaction execution and auditing: payers can finalize transactions without the payee online, and authorized auditors can recover identities and amounts from on-chain data without user cooperation. This design improves usability, reliability, and compliance in real-world blockchain deployments.

    \item \textbf{Renewable Anonymous Certificate.} We define renewable anonymous certificates (RAC) for deriving publicly verifiable certificates on randomized account addresses from certificates on registered accounts. RAC allows validators to verify that an anonymous output is derived from a certified account, without learning the underlying account identity. This prevents identity forgery and fake transactions without relying on recording auditors.

    \item \textbf{Novel Transaction Structure.} We design a novel transaction structure that enables both payees and auditors to independently access the necessary transaction data using their respective keys, without requiring any interaction between parties. This structure is compatible with blockchain environments based on elliptic curves, ensuring seamless integration with existing systems while maintaining computational efficiency.

    \item \textbf{Formal Security Proofs and Efficient Implementation.} We give a game-based security model and prove authenticity, anonymity, confidentiality, and soundness. We also provide a concrete instantiation of SilentLedger and evaluate its performance against representative auditable payment systems.
\end{enumerate}

\section{Related Work}

In recent years, privacy-preserving blockchain systems have evolved from early privacy-enhancing solutions to auditable privacy-preserving solutions. Here, we outline this progression with respect to system models and functional implementations.

\subsection{Privacy-enhancing Solutions}

While Bitcoin introduced the foundational concept of decentralized cryptocurrency, it did not originally prioritize transaction privacy, exposing all transaction details to the public. This transparency limitation prompted the development of enhanced privacy cryptocurrencies, with early solutions including Monero and Zerocash~\cite{Zerocash}. Monero employs ring signatures and temporary payment addresses to obscure participant identities, while Zerocash achieves comprehensive transaction privacy through the integration of cryptographic commitments and zk-SNARKs. Building upon Monero's architectural foundation, Sun et al.~\cite{Ringct_2.0} enhanced its anonymity capabilities with RingCT, introducing a linkable ring signature scheme based on an accumulator that provides stronger privacy guarantees.
To address the scalability limitations inherent in UTXO-based systems, Fuchsbauer et al.~\cite{Aggregate_cash_systems} proposed a novel approach utilizing address accumulation and homomorphic commitments for privacy transactions. The resulting Mimblewimble protocol provides efficient transaction data compression while maintaining transaction value confidentiality. However, a fundamental challenge persists across these approaches: they require users to share the random values of cryptographic commitments to complete transactions, which introduces the necessary interactivity and affects system reliability.

Several solutions have emerged to address this challenge. Fauzi et al.~\cite{Quisquis} proposed encrypting transaction details with the payees' public keys, thereby enabling independent verification and spending without coordination. Similarly, Non-interactive Mimblewimble~\cite{Non-interactive_mimblewimble} extended the original Mimblewimble framework by incorporating a non-interactive Diffie--Hellman key exchange, which allows payees to independently derive the random values of commitments required for future spending.

While these privacy-enhancing solutions effectively protect user data, they also create challenges for compliance auditing. By concealing transaction details, they risk enabling illicit activities such as money laundering and tax evasion, thereby limiting blockchain adoption in regulated financial environments. To address these concerns, recent research has turned to auditable blockchain systems, which incorporate selective traceability mechanisms that grant authorized entities access to specific transaction data while maintaining privacy for general users.

\subsection{Auditable Privacy-preserving Solutions}

Narula et al.~\cite{zkLedger} addressed this need with the zkLedger system, which employs OR proof technology within a ledger architecture. To address zkLedger's scalability limitations, Chatzigiannis et al.~\cite{MiniLedger} developed MiniLedger, which optimizes storage efficiency through the use of compact cryptographic proofs, while Chen et al.~\cite{PGC} designed PGC to reduce dependence on trusted authorities through the strategic integration of encryption algorithms and non-interactive zero-knowledge proofs. However, these systems share a critical architectural limitation in that they protect only transaction confidentiality without providing comprehensive privacy protection.

To achieve complete privacy protection while maintaining auditability, several approaches have emerged. These solutions typically operate through a model with dual accounts where users conduct transactions via temporary private accounts derived from their registered accounts, with auditors possessing specialized cryptographic keys to recover registered account and transaction data when necessary for compliance purposes. Representative solutions include Traceable Monero~\cite{Traceable_monero}, Platypus~\cite{Platypus}, and ACA~\cite{Aca}, which all employ the auditor's public key for identity encryption while applying different approaches for amount concealment. Specifically, Traceable Monero and Platypus use homomorphic commitments to conceal transaction amounts, whereas ACA employs homomorphic encryption under users' secret keys. However, these designs introduce an operational limitation: they require mandatory interactions during the auditing process, such as audit checks before transactions or disclosures of transaction amounts initiated by users.

PEReDi~\cite{Peredi} represents a notable attempt to eliminate interaction requirements during auditing by homomorphically encrypting both identity and amount information under the auditor's public key, enabling independent decryption of complete transaction details by the auditor. While this approach successfully removes the need for user involvement during audits, it introduces a different form of dependency that affects transaction autonomy by preventing users from independently validating encrypted transaction amounts, thus necessitating reliance on the auditor to sign transaction status updates and reintroducing interaction requirements.

\section{Preliminaries} \label{preliminaries}

This section provides the background needed for the following construction. We first recall a common UTXO style transaction model and then summarize the cryptographic primitives used later.

\subsection{Transaction Model}
\label{sec:two_two_model}

A 2-2 transaction refers to a transaction in the UTXO model with two inputs and two outputs. It consumes two existing unspent outputs, denoted by $\mathsf{In}_1$ and $\mathsf{In}_2$, and creates two new outputs, denoted by $\mathsf{Out}_1$ and $\mathsf{Out}_2$. Let $v^{\mathrm{in}}_1$ and $v^{\mathrm{in}}_2$ be the input values, and let $v^{\mathrm{out}}_1$ and $v^{\mathrm{out}}_2$ be the output values. Ignoring transaction fees, validity requires
$
v^{\mathrm{in}}_1+v^{\mathrm{in}}_2
=
v^{\mathrm{out}}_1+v^{\mathrm{out}}_2.
$
In practice, one output may be a payment and the other a change output returned to the payer. Alternatively, both outputs may be payments to two recipients. In this paper, the 2-2 model is the basic setting for the construction, security experiments, SoK design, and performance evaluation.

\subsection{Cryptographic Primitives}

\textbf{Encryption primitives.}
SilentLedger uses two encryption primitives with different roles. The public-key encryption scheme $\mathsf{PKE}=(\mathsf{PKESetup},\mathsf{PKEKGen},\mathsf{PKEEnc},\mathsf{PKEDec})$
encrypts keys required during tracing for the auditor. Given public parameters $pp$, a key pair
$(sk,pk)\leftarrow\mathsf{PKEKGen}(pp)$, and a message $m$, the ciphertext
$c\leftarrow\mathsf{PKEEnc}(pp,m,pk)$ can be decrypted only with the matching secret key: $\mathsf{PKEDec}(c,sk)=m$.

The symmetric encryption scheme
$\mathsf{SKE}=(\mathsf{SKEnc},\mathsf{SKDec})$  hides transaction data under derived keys.
We write $c\leftarrow\mathsf{SKEnc}(pp,m,xk)$, where $m$ is a message and $xk$ is the symmetric key. Corresponding decryption requires
$\mathsf{SKDec}(c,xk)=m$. 

In our construction, \textsf{PKE} enables authorized tracing, while \textsf{SKE} hides anonymous addresses and transaction amounts.

\textbf{Non-Interactive Key Exchange.}
Non-Interactive Key Exchange (\textsf{NIKE}) allows two parties to derive a shared key from one party's secret key and the other party's public key without an online key exchange. For matching key pairs $(sk_i,pk_i)$ and $(sk_j,pk_j)$, it satisfies
\[
\mathsf{NIKE}(pk_i,sk_j)=\mathsf{NIKE}(pk_j,sk_i).
\]
In SilentLedger, public keys are authenticated by registered account addresses and certificates, and \textsf{NIKE} provides the non-interactive derivation of shared keys used in transaction generation and recovery.

\textbf{Key Derivation Function.}
The key derivation function (\textsf{KDF}) derives a group-valued key from a shared scalar obtained through \textsf{NIKE}. Given a shared scalar $c$, the derived key is
$K\leftarrow\mathsf{KDF}(c)$.
In SilentLedger, this is used as a recovery key to enable authorized recovery. 

\textbf{Reversible Amount Encoding.}
The reversible function \textsf{RF} is used to encode bounded amounts. For $v\in[0,B)$, $\mathsf{RF}(pp,v)$ encodes $v$ as a group element, and
$\mathsf{RF}^{-1}(pp,\cdot)$ recovers $v$ within the bounded range.

\subsection{Randomizable Signature Primitives}
\label{sec:randomizable_primitives}

A randomizable signature primitive transforms a signed object into a fresh signed representative of the same equivalence class. Given a signed object $(X,\sigma)$ and a randomization factor $r$, let $[X]$ denote the class of objects obtained by randomizing $X$. A generic randomizable signature primitive provides 
$
\sigma\leftarrow\mathsf{Sign}(sk,X)
$
and
$
(X',\sigma')\leftarrow\mathsf{RandSign}(X,\sigma,r),
$
where $X'\in[X]$. 
Signatures on randomizable ciphertexts (SoRC) instantiate this paradigm over ciphertexts~\cite{Signatures_on_randomizable_ciphertexts,Efficient_signatures_on_randomizable_ciphertexts}, while randomizable signatures on randomizable commitments (RSoRC) extend it to commitments by endowing randomized commitments with authentication capabilities~\cite{BlindHub}. Such primitives have been used in applications including e-voting, blind signatures, anonymous credentials, and payment channel hubs~\cite{Signatures_on_randomizable_ciphertexts,Efficient_signatures_on_randomizable_ciphertexts,BlindHub}.

In this paper, SilentLedger requires a randomizable signature primitive that supports public authorization verification of identities. Although RSoRC provides authentication for commitments, its design based on commitments does not match our privacy mechanism based on encryption. Following the randomization paradigm, Section~\ref{sec:rac} introduces RAC for certified account identities.

\subsection{Proof Primitives}

\textbf{Signatures of Knowledge.}
A signature of knowledge (\textsf{SoK}) allows a prover to convince a verifier that it knows a witness $w$ for a public statement $\mathsf{stmt}$, while binding the proof to a message $m$. For an efficiently decidable relation $\mathcal{R}$, we use
$
\mathsf{SoK}\{(\mathsf{stmt},w):\mathcal{R}(\mathsf{stmt},w)=1\}(m)
$
to denote a proof of knowledge of $w$ for $\mathsf{stmt}$ with respect to message $m$. A \textsf{SoK} scheme consists of the following algorithms.

\begin{itemize}
	\item $\pi\gets\mathsf{SoK.Prove}(pp,\mathsf{stmt},w,m)$:
	Given public parameters $pp$, a statement $\mathsf{stmt}$, a witness $w$, and a message $m$, this algorithm outputs a signature of knowledge $\pi$.

	\item $0/1 \gets\mathsf{SoK.Verf}(pp,\mathsf{stmt},m,\pi)$:
	Given $pp$, $\mathsf{stmt}$, $m$, and $\pi$, this algorithm outputs $1$ if $\pi$ is valid, and $0$ otherwise.
\end{itemize}

\begin{definition}\label{def:sok-soundness}
A \textsf{SoK} scheme is \textit{sound} for relation $\mathcal{R}$ if no $\mathcal{PPT}$ adversary can produce an accepting proof for a false statement, except with negligible probability. Let $\mathcal{L}_{\mathcal{R}}=\{\mathsf{stmt}:\exists w,(\mathsf{stmt},w)\in\mathcal{R}\}$. For any $\mathcal{PPT}$ adversary $\mathcal{A}$, where $(\mathsf{stmt},m,\pi)\gets\mathcal{A}(pp)$, we have:
\begin{align*}
\Pr\left[
\begin{array}{c}
\mathsf{SoK.Verf}(pp,\mathsf{stmt},m,\pi)=1\\
\land\ \mathsf{stmt}\notin\mathcal{L}_{\mathcal{R}}
\end{array}
\right]
\leq \mathsf{negl}(\lambda).
\end{align*}
\end{definition}

\begin{definition}\label{def:sok-zk}
A \textsf{SoK} scheme is \textit{zero-knowledge} for relation $\mathcal{R}$ if there exists a $\mathcal{PPT}$ simulator $\mathcal{S}$ such that, for any $(\mathsf{stmt},w)\in\mathcal{R}$ and any message $m$, the simulated proof is computationally indistinguishable from the real proof:
\begin{align*}
\mathsf{SoK.Prove}(pp,\mathsf{stmt},w,m)
\approx_c
\mathcal{S}(pp,\mathsf{stmt},m).
\end{align*}
\end{definition}

\textbf{Range Proofs.}
A range proof allows a prover to show that a committed value lies in a valid interval without revealing the value. For a Pedersen commitment $C=vG+rH$, the range proof relation is defined as
$
\mathcal{R}_{\mathsf{rp}}=
\left\{
(C,A,B,G,H,v,r):
C=vG+rH \land v\in[A,B)
\right\}.
$
We use Bulletproofs++~\cite{Bulletproofs++}, a transparent zero-knowledge argument system that supports aggregated range proofs. For commitments $\{C_i=v_iG+r_iH\}_{i=1}^{k}$, we use the following interface.

\begin{itemize}
	\item $\pi_{\mathsf{bp}}\gets
	\mathsf{BP.Prove}
	(\{C_i,A_i,B_i\}_{i=1}^{k},G,H,\vec{g},\vec{h},
	\{(v_i,r_i)\}_{i=1}^{k})$:
	Given the commitments, public ranges, generators, and openings, this algorithm outputs an aggregated Bulletproofs++ proof showing that each $C_i$ opens to a value $v_i\in[A_i,B_i)$.

	\item $0/1 \gets
	\mathsf{BP.Verf}
	(\{C_i,A_i,B_i\}_{i=1}^{k},G,H,\vec{g},\vec{h},
	\pi_{\mathsf{bp}})$:
	Given the public inputs and the proof, this algorithm outputs $1$ if $\pi_{\mathsf{bp}}$ is valid, and $0$ otherwise.
\end{itemize}

The Bulletproofs++ range proof protocol satisfies three properties, namely \textit{completeness}, \textit{knowledge soundness}, and \textit{zero-knowledge}~\cite{Bulletproofs++}.

\section{Renewable Anonymous Certificate}
\label{sec:rac}

This section defines the syntax and security notions of RAC, gives a bilinear-group instantiation, and proves the security of the instantiation.

\subsection{Syntax and Security Definitions}

The syntax of a RAC scheme consists of the following algorithms.

\begin{itemize}
	\item $pp \gets \mathsf{Setup}(\lambda)$. Initializes public parameters from the security parameter $\lambda$.
	\item $C \gets \mathsf{CertGen}(pp)$. Generates an identity $C$.
	\item $C' \gets \mathsf{Rndmz}(C, r)$. Derives an anonymous identity $C'$ from the identity $C$ using randomness $r$.
	\item $(sk, vk) \gets \mathsf{SKeyGen}(pp)$. Generates a signing/verification key pair $(sk, vk)$.
	\item $\sigma \gets \mathsf{Sign}(sk, C)$. Produces a signature $\sigma$ on $C$. The pair $(C, \sigma)$ forms a certificate.
	\item $\sigma' \gets \mathsf{Adapt}(\sigma, r)$. Adapts the signature $\sigma$ to $\sigma'$ for the randomized identity $C'$ using randomness $r$.
	\item $0/1 \gets \mathsf{Verify}(vk, C, \sigma)$. Outputs $b\in\{0,1\}$ indicating validity of a certificate $(C, \sigma)$ (analogously for $(C', \sigma')$).
\end{itemize}

\textbf{Correctness.}
We first state the correctness for honestly generated RAC certificates. The correctness of adapted certificates is verified for the concrete construction below.

\begin{definition}\label{def:rac_correctness}
	A RAC scheme is \textit{correct} if, for all $pp \in \mathcal{PP}$, all $(sk, vk)$ in the range of $\mathsf{SKeyGen}(pp)$, and all $C \in \mathcal{C}_{pp}$:
	\begin{align*}
		\Pr[\mathsf{Verify}(vk,C,\mathsf{Sign}(sk,C))=1]=1.
	\end{align*}
\end{definition}

\begin{definition}\label{Signature-adaptation}
A scheme is signature-adaptable if, for all $pp \in \mathcal{PP}$, any $(sk,vk)$ output by $\mathsf{SKeyGen}(pp)$, any $C \in \mathcal{C}_{pp}$, and any $r \in \mathcal{R}_{pp}$, letting $\sigma \leftarrow \mathsf{Sign}(sk,C)$ and $C' \leftarrow \mathsf{Rndmz}(C,r)$, the two distributions
$\mathsf{Adapt}(\sigma,r)$ and $\mathsf{Sign}(sk, C')$ are identical.
\end{definition}

\textbf{Unforgeability.} For an identity $C$, let
$
[C]=\{C'\in\mathcal{C}_{pp}: \exists r\in\mathcal{R}_{pp}, C'=\mathsf{Rndmz}(C,r)\}
$
be its randomization class. After a signing query on $C$, all identities in $[C]$ are recorded in the query set $Q$ since a signature on $C$ can be adapted to any randomized identity in this class. The adversary wins only if it outputs a valid signature on some $C^* \notin Q$.

\begin{definition}\label{Unforgeability}
	Let $\mathbf{EUF}^{\mathcal{A}}_{\mathsf{RAC}}(\lambda)$ be the experiment defined in Fig.~\ref{unforgeability}. A RAC scheme achieves \textit{unforgeability} if, for every PPT adversary $\mathcal{A}$,
\begin{align*}
	\Pr[\mathbf{EUF}^{\mathcal{A}}_{\mathsf{RAC}}(\lambda)=1]\le \mathsf{negl}(\lambda).
\end{align*}
\end{definition}

\begin{figure}[!b]
    \centering
    \begin{tcolorbox}[
        sharp corners,
        colback=white,
        boxrule=0.4mm,
        width=\columnwidth,
        left=1.2mm,
        right=1.2mm,
        top=0mm,
        bottom=0mm
    ]
    \footnotesize
    \begin{minipage}[t]{0.60\linewidth}
        \centering
        \underline{$\mathbf{EUF}^{\mathcal{A}}_{\mathsf{RAC}}(\lambda)$}\\[0.4ex]
        \begin{tabular}{@{}l@{}}
        $Q\gets\emptyset$\\
        $pp\gets\mathsf{Setup}(1^\lambda)$\\
        $(sk,vk)\gets\mathsf{SKeyGen}(pp)$\\
        $(C^*,\sigma^*)\gets
        \mathcal{A}^{\mathcal{O}_{\mathsf{Sign}}(\cdot)}(vk)$\\
        $\mathbf{return}\ 
        \mathsf{Verify}(vk,C^*,\sigma^*)=1
        \land C^*\notin Q$
        \end{tabular}
    \end{minipage}
    \hfill
    \begin{minipage}[t]{0.34\linewidth}
        \centering
        \underline{$\mathcal{O}_{\mathsf{Sign}}(C)$}\\[0.4ex]
        \begin{tabular}{@{}l@{}}
        $\sigma\gets\mathsf{Sign}(sk,C)$\\
        $Q\gets Q\cup[C]$\\
        $\mathbf{return}\ \sigma$
        \end{tabular}
    \end{minipage}
    \end{tcolorbox}
    \caption{Unforgeability experiment for RAC.}
    \label{unforgeability}
\end{figure}

\begin{comment}
\begin{figure*}[b!]
	\begin{tcolorbox}[sharp corners, colback=white, boxrule=0.4mm]
		$\mathsf{Setup}(1^\lambda)$: Generate public parameters $pp = (g, G_1, \mathbb{G}_1, G_2, \mathbb{G}_2, \mathbb{G}_{T}) \gets \mathsf{BGGen}(\lambda)$, in which $\mathbb{G}_1 \times \mathbb{G}_2 \rightarrow \mathbb{G}_{T}$, and set $g = e(G_1, G_2)$.

		$\mathsf{CertGen}(pp)$: Randomly select $r \gets \mathbb{Z}^*_p$, compute $C = r G_1$, and return $(r, C)$.

		$\mathsf{Rndmz}(C, r')$: Randomly select $r' \gets \mathbb{Z}_p$ and return $C' = C + r' G_1$.

		$\mathsf{SKeyGen}(pp)$: Randomly select $sk := x \gets \mathbb{Z}^*_p$, compute $vk := (X =xG_2)$ and return $(sk, vk)$.

		$\mathsf{Sign}(x, C)$: Randomly select $s \gets \mathbb{Z}^*_p$, and compute:
		$Z := {s}^{-1}(G_1 + xC)$, $U := sG$, $\widehat{U} := s G_2$, $Y := {s}^{-1}xG_1$.
		Return $\sigma := (Z, U, \widehat{U}, Y)$.

		$\mathsf{Adapt}(\sigma , r')$: Randomly select $s' \gets \mathbb{Z}^*_p$, and compute:
		$Z' := {s'}^{-1}(Z + r'Y)$, $U' := s'U$, $\widehat{U}' := s'\widehat{U}$, $Y' := {s'}^{-1}Y$.
		Return $\sigma' := (Z', U', \widehat{U}', Y')$.

		$\mathsf{Verify}(X, C, \sigma)$: Return 1 if the following equations hold, 0 otherwise:
		\begin{align*}
			e(Z, \widehat{U}) = e(G_1, G_2)e(C, X) \hspace{2em}
			e(G_1, \widehat{U}) = e(U, G_2) \hspace{2em}
			e(Y,\widehat{U}) = e(G_1, X)
		\end{align*}
	\end{tcolorbox}
	\caption{Instantiation of RAC.}
	\label{RAC}
\end{figure*}
\end{comment}
\subsection{Instantiation of RAC}

We present a concrete instantiation of the RAC scheme over bilinear groups and analyze its unforgeability in the generic group model. Let $(\mathbb{G}_1,\mathbb{G}_2,\mathbb{G}_T,e)$ be bilinear groups of prime order $p$ with $e:\mathbb{G}_1\times\mathbb{G}_2\to\mathbb{G}_T$ and generators $G_1\in\mathbb{G}_1$, $G_2\in\mathbb{G}_2$.
\begin{itemize}[noitemsep]
	\item $\mathsf{Setup}(1^\lambda)$: Output $pp=(g,G_1,G_2,\mathbb{G}_1,\mathbb{G}_2,\mathbb{G}_T,e)$ and set $g:=e(G_1,G_2)$.
	\item $\mathsf{CertGen}(pp)$: Sample $r\gets\mathbb{Z}_p^{*}$, set $C:=rG_1$, and output $(r,C)$.
	\item $\mathsf{Rndmz}(C,r')$: Given $C$ and $r'\gets\mathbb{Z}_p$, return $C':=C+r'G_1$.
	\item $\mathsf{SKeyGen}(pp)$: Sample $x\gets\mathbb{Z}_p^{*}$, set $X:=xG_2$, and output $(sk:=x, vk:=X)$.
	\item $\mathsf{Sign}(sk,C)$: Given $sk=x$ and $C$, sample $s\gets\mathbb{Z}_p^{*}$ and compute $Z := {s}^{-1}(G_1 + xC)$, $U := sG_1$, $\widehat{U} := sG_2$, $Y := {s}^{-1}xG_1$, and return $\sigma := (Z, U, \widehat{U}, Y)$.
	\item $\mathsf{Adapt}(\sigma,r')$: Given $\sigma$ and $r'$, sample $s'\gets\mathbb{Z}_p^{*}$ and compute $Z' := {s'}^{-1}(Z + r'Y)$, $U' := s'U$, $\widehat{U}' := s'\widehat{U}$, $Y' := {s'}^{-1}Y$, and output $\sigma':=(Z',U',\widehat{U}',Y')$ for the randomized identity $C':=C+r'G_1$.
	\item $\mathsf{Verify}(vk,C,\sigma)$: On input $(X,C,\sigma)$, return 1 if:
			$e(Z, \widehat{U}) = e(G_1, G_2)e(C, X)$,
			$e(G_1, \widehat{U}) = e(U, G_2)$, and
			$e(Y,\widehat{U}) = e(G_1, X)$.
	Verification for $(C',\sigma')$ is identical.
\end{itemize}

A key feature of our scheme is that any holder of a valid certificate $(C,\sigma)$ can independently choose $r'$ and derive an anonymous certificate $(C',\sigma')$ without learning either the signing key $x$ or the original certificate randomness $r$. Both certificates are validated by the same $\mathsf{Verify}$ procedure.

\textit{Correctness.}
For an honestly generated certificate $(C,\sigma)$ with $C=rG_1$ and
$\sigma=(Z,U,\widehat{U},Y)$, where
$Z=s^{-1}(G_1+xC)$, $U=sG_1$, $\widehat{U}=sG_2$, and $Y=s^{-1}xG_1$, bilinearity gives
	\begin{align*}
	e(Z, \widehat{U}) &= e(s^{-1}(G_1 + xC), sG_2) = e(G_1 + xC, G_2)\\
	&= e(G_1, G_2)e(C, xG_2) = e(G_1, G_2)e(C, X),\\
	e(G_1, \widehat{U}) &= e(G_1, sG_2) = e(sG_1, G_2) = e(U, G_2),\\
	e(Y, \widehat{U}) &= e(s^{-1}xG_1, sG_2)  = e(G_1, xG_2) = e(G_1, X).
	\end{align*}
Thus $\mathsf{Verify}(X,C,\sigma)=1$. Since $\mathsf{Adapt}$ preserves the same verification equations for $C'=C+r'G_1$, the adapted certificate also verifies.

\begin{comment}
In SilentLedger, the account address $S$ serves as the identity $C$, the transaction shared random value $c$ functions as the randomness parameter $r$, and the anonymous address $Q$ acts as the derived pseudonym $C'$. This design equivalence enables seamless integration of RAC into our transaction system and limits authentication overhead during transaction processing to the cost of executing the \textsf{Adapt} algorithm.
\end{comment}

\subsection{Security Analysis of RAC}
\label{sec:rac_security_analysis}

We next analyze the concrete RAC scheme with respect to \textit{signature-adaptability} and \textit{unforgeability}.

\begin{theorem}
	The RAC scheme is signature-adaptable under maliciously generated keys.
\end{theorem}

\begin{IEEEproof}[Proof]
	Let $C=rG_1$ and let $\sigma=(Z,U,\widehat{U},Y)$ be any valid signature, written as $Z=zG_1$, $U=sG_1$, $\widehat{U}=\widehat{s}G_2$, and $Y=tG_1$.
	The second verification equation gives $s=\widehat{s}$, and the remaining equations imply
$
zs=1+rx, ts=x.
$

	Now sample $r'\in\mathbb{Z}_p$, and for $C'=C+r'G_1$, any valid signature $\sigma'=(Z',U',\widehat{U}',Y')=(z'G_1,s'G_1,s'G_2,t'G_1)$ satisfies
	\[
		z's' = 1 + (r + r')x = zs + r'x = zs + r'ts,~t's' = x = ts.
	\]
	Since $s,s'\ne 0$, define $\rho:=s'/s\in\mathbb{Z}_p^*$. Then the above equations imply
\begin{align*}
	Z' = \rho^{-1}(Z+r'Y),
	Y' = \rho^{-1}Y,
	U' = \rho U,
	\widehat{U}' = \rho\widehat{U}.
\end{align*}
	Thus, for every valid signature $\sigma'$ on $C'=C+r'G_1$, there exists a unique $\rho\in\mathbb{Z}_p^*$ such that
\[
\sigma'=\left(\rho^{-1}(Z+r'Y),\rho U,\rho\widehat{U},\rho^{-1}Y\right).
\]
	Conversely, every $\rho\in\mathbb{Z}_p^*$ defines a valid signature on $C'$. Since $\mathsf{Adapt}$ samples $\rho$ uniformly from $\mathbb{Z}_p^*$, its output is uniformly distributed over the set of valid signatures on $C'$, as required. 
\end{IEEEproof}

	We prove unforgeability in the generic group model~\cite{GGM} for Type-3 bilinear groups. The adversary sees only random encodings and accesses group operations and pairings through generic oracles. Here is a proof sketch. 

\begin{theorem}
	A generic adversary $\mathcal{A}$ making at most $q$ generic oracle queries and at most $k$ signing queries wins $\mathbf{EUF}^{\mathcal{A}}_{RAC}(\lambda)$ with probability at most
	$
	\frac{(4k+2)(q+4k+3)^2}{2(p-1)}.
	$
	In particular, when $p>2^\lambda$, this probability is negligible in $\lambda$.
\end{theorem}
\begin{IEEEproof}[Proof Sketch]
Fix a generic adversary $\mathcal{A}$ that makes at most $q$ generic oracle queries and at most $k$ signing queries. We compare the real generic group experiment with an ideal symbolic experiment. In the symbolic experiment, each group handle is represented by a formal expression determined by the hidden signing key, the signing randomness used in oracle responses, and the public operations requested by $\mathcal{A}$. In this experiment, $\mathcal{A}$ can only combine handles through group operations and pairings.

We first consider the symbolic experiment. When the signing oracle answers a query on $C_i$, it records the whole randomization class $[C_i]$, since any certificate on $C_i$ can be adapted to any randomized identity in this class. Suppose that $\mathcal{A}$ outputs an accepting certificate on an identity $C^*$. The RAC verification equations force the certificate components and the signed identity to be algebraically consistent with the same signing key and one signing randomness value. In the symbolic experiment, the only way to satisfy these constraints is to derive the forged certificate from one previous signing query. Hence there must exist an index $i_0$ and a public scalar $\rho\in\mathbb{Z}_p$ such that
$
C^*=C_{i_0}+\rho G_1.
$
Thus $C^*$ belongs to $[C_{i_0}]$, which has already been recorded by the signing oracle. Therefore, the adversary cannot produce a fresh valid certificate in the symbolic experiment.

It remains to account for the difference between the symbolic experiment and the real generic group experiment. In the symbolic experiment, two different formal expressions remain distinct unless their equality follows from the algebraic rules of the game. In the real experiment, however, these expressions are evaluated using random hidden values and represented by concrete random encodings. The two experiments can differ only if two formally distinct expressions accidentally evaluate to the same group element. Such an event is an encoding collision rather than a structural forgery.

The full proof shows that every possible collision is governed by a nontrivial algebraic relation of degree at most $4k+2$ in the hidden random values. Since the experiment creates at most $q+4k+3$ handles, applying the Schwartz--Zippel lemma to each possible collision and taking a union bound over all pairs of handles gives
\[
\Pr[E_{\mathsf{coll}}]
\le
\frac{(4k+2)(q+4k+3)^2}{2(p-1)}.
\]
Because the adversary cannot win in the symbolic experiment, its winning probability in the real generic group experiment is bounded by this collision probability, which is negligible when $p>2^\lambda$.
\end{IEEEproof}

\section{SilentLedger}

In this section, we describe the construction of SilentLedger under the 2-2 transaction model defined in Section~\ref{sec:two_two_model} and prove its security guarantees. For readability, Table~\ref{tab:notation_silentledger} summarizes the key variables used below.

\begin{table}[!t]
	\caption{Notation reference.}
	\label{tab:notation_silentledger}
	\centering
	\normalsize
	\setlength{\tabcolsep}{1.8pt}
	\renewcommand{\arraystretch}{1.03}
	\begin{tabularx}{\columnwidth}{@{}>{\centering\arraybackslash}m{0.12\columnwidth}>{\centering\arraybackslash}m{0.2\columnwidth}>{\raggedright\arraybackslash}X@{}}
		\toprule
		\textbf{Visibility} & \textbf{Notation} & \textbf{Meaning} \\
		\midrule

		\multirow{13}{*}{Public}
		& $\lambda$ & Security parameter. \\
		& $pp$ & Public parameters. \\
		& $i$ & Account or output index. \\
		& $T$ & Tracing public key. \\
		& $X$ & RAC verification key. \\
		& $Acc_i$ & Anonymous account. \\
		& $Q_i$ & Anonymous address. \\
		& $cm_i$ & Encrypted amount. \\
		& $R_i$ & Ephemeral public key. \\
		& $CK_i$ & Ciphertext for the recovery key. \\
		& $\sigma'_i$ & Adapted RAC signature. \\
		& $(\mathsf{stmt},m,\pi)$ & Statement, message, and proof of \textsf{SoK}. \\
		& $tx$ & Posted transaction. \\

		\midrule
		\multirow{6}{*}{Payer}
		& $sk_i$ & Account secret key. \\
		& $(vk_i,V_i)$ & Viewing key pair. \\
		& $\sigma_i$ & RAC signature for account $i$. \\
		& $r_i$ & Ephemeral secret key. \\
		& $c_i$ & Access key. \\
		& $w$ & Witness for \textsf{SoK}. \\

		\midrule
		\multirow{2}{*}{Auditor}
		& $mk$ & Tracing secret key. \\
		& $x$ & RAC signing secret key. \\

		\midrule
		\multirow{5}{*}{\makecell[c]{Payer,\\Auditor}}
		& $S_i$ & Account address. \\
		& $K_i$ & Recovery key. \\
		& $v_i$ & Transaction amount. \\
		& $vx_i$ & Encoded amount. \\
		& $tc_i$ & Symmetric key for amount recovery. \\

		\midrule
		\multicolumn{3}{@{}p{\columnwidth}@{}}{\textit{Note:} Public denotes public parameters or on-chain data. In a transaction, unhatted variables denote input variables of the payer, while $\widehat{\cdot}$ denotes output variables of the payee. Data intended for the auditor, such as $CK_i$ and $R_i$, remains unhatted.}\\
	\end{tabularx}
\end{table}

\subsection{Construction of SilentLedger} \label{sec:silentledger_construction}

SilentLedger consists of seven algorithms: \textsf{Setup}, \textsf{MKGen}, \textsf{UKGen}, \textsf{AAGen}, \textsf{Trans}, \textsf{VerfTX}, and \textsf{Trace}. The RAC scheme serves as the authorization layer: \textsf{CertGen} generates account addresses, \textsf{Sign} issues certificates, and \textsf{Adapt} derives verifiable anonymous certificates for transaction outputs.

\begin{itemize}
	\item
	$pp \gets \mathsf{Setup}(\lambda)$:
	This algorithm initializes the system with security parameter $\lambda$ and generates the public parameters $pp$ for all underlying primitives.
	\item
	$((mk,T),(x,X)) \gets \mathsf{MKGen}(pp)$:
		This algorithm is executed by the auditor during system initialization. It generates a tracing key pair $(mk,T)$ using \textsf{PKEKGen}, and a signing/verification key pair $(x,X)$ using \textsf{RAC.SKeyGen}.
	\item
	$((sk_i,S_i),(vk_i,V_i),\sigma_i) \gets \mathsf{UKGen}(pp)$:
		This algorithm generates keys and credentials for users. It first obtains an account key pair $(sk_i,S_i)$ through \textsf{PKEKGen}, where the account address $S_i$ serves as the certified identity input to \textsf{RAC}. The auditor then issues a signature $\sigma_i\leftarrow\mathsf{RAC.Sign}(x,S_i)$. The user also generates a viewing key pair $(vk_i,V_i)$ using \textsf{PKEKGen}. The public components $(S_i,V_i,\sigma_i)$ are used as registration information for receiving future transactions, while $(sk_i,vk_i)$ remain local to the user.

	\item
	$(\widehat{Acc}_i,CK_i,R_i,\widehat{\sigma}'_i) \gets \mathsf{AAGen}(pp,\widehat{v}_i,\widehat{S}_i,\widehat{\sigma}_i,\widehat{V}_i,T)$:
	This algorithm generates an anonymous account for the payee with account address $\widehat{S}_i$ and viewing public key $\widehat{V}_i$. 	It first generates an ephemeral key pair $(r_i,R_i)$ and derives the access key for the payee as
	$\widehat{c}_i\leftarrow\mathsf{NIKE}(\widehat{V}_i,r_i)$.
	It then derives a recovery key $\widehat{K}_i\leftarrow\mathsf{KDF}(\widehat{c}_i)$
	and encrypts $\widehat{K}_i$ for the auditor as $CK_i\leftarrow\mathsf{PKEEnc}(pp,\widehat{K}_i,T)$ with randomness.

	To hide the payee's identity, the algorithm derives an anonymous address $\widehat{Q}_i\leftarrow\mathsf{RAC.Rndmz}(\widehat{S}_i,\widehat{c}_i)$, equivalently $\mathsf{SKEnc}(pp,\widehat{S}_i,\widehat{K}_i)$, and an adapted signature $\widehat{\sigma}'_i\leftarrow\mathsf{Adapt}(\widehat{\sigma}_i,\widehat{c}_i)$ that remains publicly verifiable for $\widehat{Q}_i$. 
	To hide the amount, it encodes $\widehat{v}_i$ as $\widehat{vx}_i\leftarrow\mathsf{RF}(pp,\widehat{v}_i)$, derives the symmetric key $\widehat{tc}_i\leftarrow \mathsf{NIKE}(T,\widehat{c}_i)$, and outputs the encrypted amount $\widehat{cm}_i\leftarrow\mathsf{SKEnc}(pp,\widehat{vx}_i,\widehat{tc}_i)$. The anonymous account is defined as $\widehat{Acc}_i=(\widehat{Q}_i,\widehat{cm}_i)$. Thus, the payer can create a spendable and traceable output using only public registration data.

	\item
	$tx \gets \mathsf{Trans}(pp,Acc_1,Acc_2,\widehat{Acc}_1,\widehat{Acc}_2,T,w)$:
Under the 2-2 transaction model, payer $P_0$ spends two input accounts $Acc_1,Acc_2$ and creates two output accounts $\widehat{Acc}_1,\widehat{Acc}_2$. For each output, the payer invokes \textsf{AAGen} and obtains $(\widehat{Acc}_i,CK_i,R_i,\widehat{\sigma}'_i)$. The payer then constructs the public statement $\mathsf{stmt}$ for the transaction relation $\mathcal{L}_{tx}$ and the corresponding witness $w$. Let $m$ denote the transaction message bound to the proof, which contains the public transaction data generated above. The payer generates a non-interactive proof
$
    \pi \gets \mathsf{SoK.Prove}(pp,\mathsf{stmt},w,m).
$
The relation $\mathcal{L}_{tx}$ requires that (i) the input accounts are spendable, (ii) the output accounts are correctly generated, (iii) each output is bound to a certified account address, (iv) the recovery keys $\widehat{K}_i$ are consistent with the access keys $\widehat{c}_i$, (v) the balance equation holds, and (vi) all transaction amounts lie in the valid range. Its concrete decomposition into $\mathcal{L}_{tx_1}$ and $\mathcal{L}_{tx_2}$ is given in Section~\ref{Instantiation}. The transaction is posted as $tx=(\mathsf{stmt},m,\pi)$, where $\mathsf{stmt}$ collects the public statement of $\mathcal{L}_{tx}$, $m$ is the transaction message bound to the proof, and $\pi$ is the non-interactive \textsf{SoK} proof for $\mathsf{stmt}$ with respect to $m$. Thus, validation requires only public data and no payee or auditor participation.

\item
$\{0,1\} \gets \mathsf{VerfTX}(pp,tx, X)$:
This public algorithm verifies the \textsf{SoK} proof and checks each adapted certificate $\widehat{\sigma}'_i$ for $\widehat{Q}_i$ contained in $tx$ via \textsf{SoK.Verf} and \textsf{RAC.Verify}. It accepts only if all checks pass. The verification uses only public parameters, posted transaction data, and public verification keys.

	\item
	$(\widehat{S}_i,\widehat{v}_i) \gets \mathsf{Trace}(pp,mk,\widehat{Acc}_i,CK_i,R_i)$:
		This tracing algorithm is executed by the auditor using its tracing secret key $mk$ and public on-chain data. The auditor decrypts $CK_i$ to recover the recovery key $\widehat{K}_i\leftarrow\mathsf{PKEDec}(CK_i,mk)$. It then recovers the payee's account address as $\widehat{S}_i\leftarrow\mathsf{SKDec}(\widehat{Q}_i,\widehat{K}_i)$. To recover the amount, the auditor derives $\widehat{tc}_i\leftarrow \mathsf{NIKE}(mk, R_i)$, decrypts $\widehat{vx}_i\leftarrow\mathsf{SKDec}(\widehat{cm}_i,\widehat{tc}_i)$, and obtains $\widehat{v}_i\leftarrow\mathsf{RF}^{-1}(pp,\widehat{vx}_i)$. 
\end{itemize}

\textbf{Completeness.}
For any honestly generated transaction $tx\gets\mathsf{Trans}(pp,Acc_1,Acc_2,\widehat{Acc}_1,\widehat{Acc}_2,T,w)$, the correctness of \textsf{SoK}, RAC verification, \textsf{NIKE}, \textsf{KDF}, \textsf{SKEnc}/\textsf{SKDec}, and \textsf{RF} ensures public verification and authorized tracing:
\begingroup
\setlength{\abovedisplayskip}{3pt}
\setlength{\belowdisplayskip}{3pt}
\setlength{\abovedisplayshortskip}{2pt}
\setlength{\belowdisplayshortskip}{2pt}
\setlength{\jot}{1pt}
\small
\begin{align*}
		&\mathsf{VerfTX}(pp, \mathsf{Trans}(pp, Acc_i, \widehat{Acc}_i, T, \pi), X) \\
		&= \mathsf{SoK.Verf}(pp, \mathsf{stmt}, m, \mathsf{SoK.Prove}(pp, \mathsf{stmt}, w, m)) \land \mathsf{Verify}(X,\widehat{Q}_i,\widehat{\sigma}'_i) \\
		&=1,\\
		&\mathsf{SKDec}(\mathsf{SKEnc}(pp,\widehat{S}_i, \mathsf{KDF}(\mathsf{NIKE}(\widehat{V}_i,r_i))),\mathsf{KDF}(\mathsf{NIKE}(\widehat{V}_i,r_i)))\\
		&=\mathsf{SKDec}(\mathsf{SKEnc}(pp,\widehat{S}_i, \mathsf{KDF}(\widehat{c}_i)),\mathsf{KDF}(\widehat{c}_i))\\
		&= \mathsf{SKDec}(\mathsf{SKEnc}(pp,\widehat{S}_i, \widehat{K}_i), \widehat{K}_i)\\
		&= \mathsf{SKDec}(\widehat{Q}_i, \widehat{K}_i) \\
        &= \widehat{S}_{i},\\
        &\mathsf{RF}^{-1}(pp, \mathsf{SKDec}(\mathsf{SKEnc}(pp,\widehat{vx}_i,\widehat{tc}_i),\widehat{tc}_i))\\
		&=\mathsf{RF}^{-1}(pp, \mathsf{SKDec}(\widehat{cm}_i,\widehat{tc}_i))\\
		&= \mathsf{RF}^{-1}(pp,\widehat{vx}_i)\\
		&= \widehat{v}_i.\\
\end{align*}
\endgroup

\subsection{Security Model and Proofs}

\textbf{Threat model.} We consider users, public validators, and an authorized auditor. The auditor is trusted for certification and tracing: it holds secret keys $(mk,x)$ and may recover identities and amounts from posted transactions. Thus, anonymity and confidentiality are defined against public observers, validators, unauthorized parties, and corrupted users without the auditor's secret keys. These adversaries observe public parameters, directory records, on-chain transactions, and may corrupt user accounts through the oracles below. The auditor does not collude with them and does not participate in transaction generation or verification.

Under this threat model, we formalize the security properties through a sequence of experiments between an adversary $\mathcal{A}$ and a challenger $\mathcal{C}$. The experiments are instantiated under the 2-2 transaction model defined in Section~\ref{sec:two_two_model}.
The challenger maintains the following lists: honest/corrupted account addresses $L_{hS}/L_{cS}$, honest/corrupted anonymous accounts $L_{hQ}/L_{cQ}$, and valid transactions $L_{tx}$. A tuple $((sk_i,S_i),(vk_i,V_i),\sigma_i)\in L_{hS}$ records an honestly registered account address, viewing public key, and certificate, while $(c_i,Acc_i,v_i)\in L_{hQ}$ means that $Acc_i$ is an honestly generated anonymous account with hidden amount $v_i$ and access key $c_i$. All lists are initialized as empty. The adversary $\mathcal{A}$ may interact with the following oracles:

\begin{itemize}[noitemsep,leftmargin=*]
	\item $\mathcal{O}_{reg}$:
	$\mathcal{A}$ queries for the $i$-th new account address. $\mathcal{C}$ generates $((sk_i,S_i),(vk_i, V_i),\sigma_i) \gets \mathsf{UKGen}(pp)$, authorizes $S_i$ as the account address associated with viewing public key $V_i$, updates $L_{hS} \gets L_{hS} \cup \{((sk_i,S_i),(vk_i, V_i),\sigma_i)\}$, and returns $(S_i, V_i,\sigma_i)$ to $\mathcal{A}$.

	\item $\mathcal{O}_{aag}$:
	$\mathcal{A}$ queries an honestly generated anonymous account for a registered account address $S_i$ with viewing public key $V_i$ and amount $v_i$. If $((sk_i,S_i),(vk_i,V_i),\sigma_i)\in L_{hS}$, $\mathcal{C}$ runs \textsf{AAGen} and denotes the anonymous account by $(Acc_i,CK_i,R_i,\sigma_i')$. It records the corresponding witness $(c_i,Acc_i,v_i) \in L_{hQ}$ and returns the public values $(Acc_i,CK_i,R_i,\sigma_i')$ to $\mathcal{A}$.

	\item $\mathcal{O}_{tx}$:
	$\mathcal{A}$ queries transaction oracle with $(Acc_1,\allowbreak Acc_2,\allowbreak \widehat{S}_1,\allowbreak \widehat{V}_1,\allowbreak \widehat{\sigma}_1,\allowbreak \widehat{v}_1,\allowbreak \widehat{S}_2,\allowbreak \widehat{V}_2,\allowbreak \widehat{\sigma}_2,\allowbreak \widehat{v}_2)$. If $(c_1,Acc_1,v_1),(c_2,Acc_2,v_2)\in L_{hQ}$, the payee public triples $(\widehat{S}_i,\widehat{V}_i,\widehat{\sigma}_i)$ are registered for $i\in\{1,2\}$, and $v_1+v_2=\widehat{v}_1+\widehat{v}_2$, $\mathcal{C}$ runs
	$(\widehat{Acc}_i,CK_i,R_i,\widehat{\sigma}'_i)\gets\mathsf{AAGen}(pp,\widehat{v}_i,\widehat{S}_i,\widehat{\sigma}_i,\widehat{V}_i,T)$
	for $i\in\{1,2\}$, invokes \textsf{Trans} to obtain $tx$, updates $L_{tx}\gets L_{tx}\cup\{tx\}$, and returns $tx$ to $\mathcal{A}$.

	\item $\mathcal{O}_{corS}$:
	$\mathcal{A}$ queries to corrupt an account address $S_i$ associated with $V_i$. If $((sk_i, S_i), (vk_i, V_i),\sigma_i) \in L_{hS}$, $\mathcal{C}$ updates $L_{hS} \gets L_{hS} \setminus \{((sk_i, S_i), (vk_i, V_i),\sigma_i)\}$, $L_{cS} \gets L_{cS} \cup \{((sk_i, S_i), (vk_i, V_i),\sigma_i)\}$, returns $(sk_i, vk_i)$ to $\mathcal{A}$.

	\item $\mathcal{O}_{corQ}$:
	$\mathcal{A}$ queries to corrupt an anonymous account $Acc_i$ together with its ciphertext $CK_i$. If $(c_i, Acc_i, v_i) \in L_{hQ}$, $\mathcal{C}$ updates $L_{hQ} \gets L_{hQ} \setminus \{(c_i, Acc_i, v_i)\}$, $L_{cQ} \gets L_{cQ} \cup \{(c_i, Acc_i, v_i)\}$, returns $(c_i, v_i)$ to $\mathcal{A}$.

	\item $\mathcal{O}_{inject}$:
	$\mathcal{A}$ uploads a transaction $tx$ with $(Acc_1, v_1,\allowbreak c_1),\allowbreak (Acc_2,\allowbreak v_2, c_2) \in L_{cQ}$. If $\mathsf{VerfTX}(pp, tx,X) = 1$,
	$\mathcal{C}$ updates $L_{tx} \gets L_{tx} \cup \{tx\}$.
\end{itemize}

In the following experiments, $\mathcal{A}$ may continue to access the above oracles before and after the challenge, except that it cannot corrupt the challenged account addresses, anonymous accounts, or output accounts. The auditor's secret keys $(mk,x)$ are never given to $\mathcal{A}$ in the anonymity and confidentiality experiments. All challenge transactions must satisfy the 2-2 balance condition and the range constraints. Let $\Pi=\mathsf{(Setup,UKGen,AAGen,Trans,VerfTX,Trace)}$ denote the SilentLedger protocol.

\begin{definition} \label{def:authenticity}
	$\Pi$ achieves \textit{authenticity} if no $\mathcal{PPT}$ adversary can output a new valid transaction spending honest anonymous accounts, except with negligible probability.
\begin{align*}
	\resizebox{\columnwidth}{!}{$
		\mathrm{Pr}
		\left[
		\begin{array}{l}
			pp \gets \mathsf{Setup}(\lambda),\\
			((S,V,\sigma),(Acc_1,Acc_2),tx) \gets \mathcal{A}^{\mathcal{O}}(pp):\\
			\exists sk,vk: ((sk,S),(vk,V),\sigma) \in L_{hS} \land
			\bigwedge_{i\in\{1,2\}} \exists c_i,v_i:
			(c_i,Acc_i,v_i)\in L_{hQ} \land\\
			\mathsf{VerfTX}(pp,tx, X)=1 \land tx \notin L_{tx}
		\end{array}
		\right]$}
		\\
		\leq \mathsf{negl}(\lambda).
\end{align*}
\end{definition}

\begin{definition} \label{def:anonymity}
	$\Pi$ achieves \textit{anonymity} if public observers and unauthorized parties cannot distinguish which of two honestly generated account pairs is used in the challenge transaction, except with negligible advantage.
\begin{align*}
	\resizebox{\columnwidth}{!}{$
		\left|
		\mathrm{Pr}
		\left[
		\begin{array}{l}
			pp \gets \mathsf{Setup}(\lambda),\ (mk,T,x,X)\gets\mathsf{MKGen}(pp),\\
			(\mathsf{st},(Acc^0_1,Acc^0_2),(Acc^1_1,Acc^1_2),
			\widehat{Acc}_1,\widehat{Acc}_2)
			\gets \mathcal{A}^{\mathcal{O}}_1(pp,T,X),\\
			\text{let }(c^d_i,Acc^d_i,v^d_i)\in L_{hQ}
			\text{ for all } d\in\{0,1\}, i\in\{1,2\},\\
			b \in \{0,1\},\\
			tx \gets \mathsf{Trans}
			(pp,Acc^b_1,Acc^b_2,\widehat{Acc}_1,\widehat{Acc}_2,T,w_b),\\
			b' \gets \mathcal{A}^{\mathcal{O}}_2(\mathsf{st},tx):\\
			b=b' \land v^0_1+v^0_2=v^1_1+v^1_2
		\end{array}
		\right]
		-\frac{1}{2}
		\right|
		$}
	\\
	\leq \mathsf{negl}(\lambda).
\end{align*}
\end{definition}

\begin{definition} \label{def:confidentiality}
	$\Pi$ achieves \textit{confidentiality} if public observers and unauthorized parties cannot distinguish which of two admissible output amounts is embedded in the challenge transaction, except with negligible advantage.
\begin{align*}
	\resizebox{\columnwidth}{!}{$
		\left|
		\mathrm{Pr}
		\left[
		\begin{array}{l}
			pp \gets \mathsf{Setup}(\lambda),\ (mk,T,x,X)\gets\mathsf{MKGen}(pp),\\
			(\mathsf{st},(\widehat{S}_1,\widehat{V}_1,\widehat{\sigma}_1),(\widehat{S}_2,\widehat{V}_2,\widehat{\sigma}_2),
			(Acc_1,Acc_2),\widehat{v}^0_1,\widehat{v}^1_1)
			\gets \mathcal{A}^{\mathcal{O}}_1(pp,T,X),\\
			\text{let }(c_i,Acc_i,v_i)\in L_{hQ}
			\text{ for } i\in\{1,2\},\\
			b \in \{0,1\},\quad
			\widehat{v}^b_2 \gets v_1+v_2-\widehat{v}^b_1,\\
			(\widehat{Acc}_1,CK_1,R_1,\widehat{\sigma}'_1)
			\gets \mathsf{AAGen}(pp,\widehat{v}^b_1,\widehat{S}_1,\widehat{\sigma}_1,\widehat{V}_1,T),\\
			(\widehat{Acc}_2,CK_2,R_2,\widehat{\sigma}'_2)
			\gets \mathsf{AAGen}(pp,\widehat{v}^b_2,\widehat{S}_2,\widehat{\sigma}_2,\widehat{V}_2,T),\\
			tx \gets \mathsf{Trans}
			(pp,Acc_1,Acc_2,\widehat{Acc}_1,\widehat{Acc}_2,T,w_b),\\
			b' \gets \mathcal{A}^{\mathcal{O}}_2(\mathsf{st},tx):\\
			b=b' \land
			\bigwedge_{d\in\{0,1\}}
			\left(\widehat{v}^d_1\in[0,u^\alpha)
			\land v_1+v_2-\widehat{v}^d_1\in[0,u^\alpha)\right)
		\end{array}
		\right]
		-\frac{1}{2}
		\right|
		$}
	\\
	\leq \mathsf{negl}(\lambda).
\end{align*}
\end{definition}

\begin{definition} \label{def:soundness}
	$\Pi$ achieves \textit{soundness} if no $\mathcal{PPT}$ adversary can make public verification accept a transaction whose statement-witness relation is invalid, except with negligible probability.
\begin{align*}
	\resizebox{\columnwidth}{!}{$
		\mathrm{Pr}
		\left[
		\begin{array}{l}
			pp \gets \mathsf{Setup}(\lambda),\ (mk,T,x,X)\gets\mathsf{MKGen}(pp),\\
			tx=(\mathsf{stmt},m,\pi) \gets \mathcal{A}^{\mathcal{O}}(pp,T,X):\\
			\mathsf{VerfTX}(pp,tx,X)=1 \land
			\neg \exists w:\ (\mathsf{stmt},w)\in \mathcal{L}_{tx}
		\end{array}
		\right]
		\leq \mathsf{negl}(\lambda)
		$}.
\end{align*}
\end{definition}

We analyze SilentLedger with respect to the underlying primitives \textsf{PKE}, \textsf{SKE}, \textsf{NIKE}, \textsf{KDF}, \textsf{SoK}, and \textsf{RAC}, together with the reversible amount encoding \textsf{RF}. Authenticity captures the inability to spend honest anonymous accounts without authorization. Anonymity and confidentiality are defined against public observers and unauthorized parties. Soundness requires that no invalid transaction statement can pass public verification except with negligible probability.

For conciseness, we provide proof sketches that focus on the essential game transitions and reductions to the underlying primitives. The full adversarial experiments and oracle restrictions are specified in Definitions~\ref{def:authenticity}--\ref{def:soundness}.

\begin{theorem} \label{security}
	Assuming the security of \textsf{PKE}, \textsf{SKE}, \textsf{NIKE}, \textsf{KDF}, \textsf{SoK}, and \textsf{RAC}, and the correctness and injectivity of \textsf{RF} over the bounded amount range, SilentLedger achieves authenticity, anonymity, confidentiality, and soundness.
\end{theorem}

\begin{IEEEproof}[Proof] \label{theorem_proof}
	We prove the theorem by establishing the four properties defined above.
\begin{lemma} \label{lem:authenticity}
	Assuming the unforgeability of \textsf{RAC} and the EUF-CMA security and soundness of \textsf{SoK}, SilentLedger achieves authenticity.
\end{lemma}

\begin{IEEEproof}[Proof]
	Let $E_0$ be the event that $\mathcal{A}$ wins the authenticity experiment. In Game~0, the challenger answers all oracles as defined and records every transaction generated through $\mathcal{O}_{tx}$ in $L_{tx}$.

	In Game~1, the challenger guesses the honest account address and the two honest anonymous accounts that will appear in the final forgery. If the adversary corrupts one of these guessed accounts, or if its final output does not use these accounts, the challenger aborts. Let $W_1$ be the non-abort event. Since the guessed accounts are chosen uniformly from the honest lists, we have
	\begin{align*}
		\Pr[W_1]
		\geq
		\frac{1}{|L_{hS}|\cdot |L_{hQ}|\cdot (|L_{hQ}|-1)}.
	\end{align*}
	Conditioned on $W_1$, the view of $\mathcal{A}$ is identical to that in Game~0. Hence,
	\begin{align*}
		\Pr[E_1]
		\geq
		\Pr[E_0]\cdot
		\frac{1}{|L_{hS}|\cdot |L_{hQ}|\cdot (|L_{hQ}|-1)}.
	\end{align*}

	In Game~2, the challenger replaces the \textsf{SoK} proofs returned by $\mathcal{O}_{tx}$ with simulated proofs for the same public statements. By the zero-knowledge property of \textsf{SoK},
	\begin{align*}
		\left|\Pr[E_2]-\Pr[E_1]\right|
		\leq
		\mathsf{negl}(\lambda).
	\end{align*}

	We now argue that $\Pr[E_2]$ is negligible. If $\mathcal{A}$ outputs a transaction $tx\notin L_{tx}$ that spends honest anonymous accounts and passes $\mathsf{VerfTX}$, then the transaction contains a fresh accepting proof for a statement that was not produced by $\mathcal{O}_{tx}$. If the statement is false, this violates the soundness of \textsf{SoK}. If the statement is true but the proof is fresh and not generated by the oracle, this yields a forgery against the EUF-CMA security of \textsf{SoK}. Moreover, if the transaction contains an adapted output certificate $\widehat{\sigma}'_i$ that verifies for an uncertified or incorrectly randomized output address $\widehat{Q}_i$, then one can construct a reduction breaking the unforgeability of \textsf{RAC}. Therefore,
	\begin{align*}
		\Pr[E_2]\leq \mathsf{negl}(\lambda).
	\end{align*}
	Combining the above inequalities, $\Pr[E_0]\leq \mathsf{negl}(\lambda)$, and authenticity follows.
\end{IEEEproof}

\begin{lemma} \label{lem:anonymity}
	Assuming the anonymity of \textsf{RAC}, the IND-CPA security of the encryption primitives, and the zero-knowledge property of \textsf{SoK}, SilentLedger achieves anonymity.
\end{lemma}

\begin{IEEEproof}[Proof]
	Let $E_0$ be the event that $\mathcal{A}$ wins the anonymity experiment. In Game~0, $\mathcal{A}$ interacts with the oracle set $\mathcal{O}$ and submits two candidate input account pairs $(Acc^0_1,Acc^0_2)$ and $(Acc^1_1,Acc^1_2)$. Both pairs must be honest anonymous accounts in $L_{hQ}$ and must satisfy the same sum of input values. The challenger samples $b\in\{0,1\}$ and generates the challenge transaction using $(Acc^b_1,Acc^b_2)$.

	In Game~1, the challenger replaces the \textsf{SoK} proofs in the challenge transaction and in transactions returned by $\mathcal{O}_{tx}$ with simulated proofs for the same public statements. Since the simulator produces proofs indistinguishable from honestly generated proofs, the zero-knowledge property of \textsf{SoK} gives
	\begin{align*}
		\left|\Pr[E_1]-\Pr[E_0]\right|
		\leq
		\mathsf{negl}(\lambda).
	\end{align*}

	In Game~2, the challenger replaces the anonymous input addresses $Q^b_i$ that depend on the branch in the challenge transaction with values corresponding to the other candidate branch, while keeping the public statement well formed. If $\mathcal{A}$ distinguishes Game~2 from Game~1 with non-negligible probability, then this distinguisher can be used to break the anonymity of \textsf{RAC}, because the only changed information is the randomized representation of the certified account addresses. Hence,
	\begin{align*}
		\left|\Pr[E_2]-\Pr[E_1]\right|
		\leq
		\mathsf{negl}(\lambda).
	\end{align*}

	In Game~3, the challenger replaces the encrypted components $cm^b_i$ for input amounts that depend on the branch in the challenge transaction with ciphertexts generated for the other branch. The adversary does not receive the auditor's secret keys $(mk,x)$ and cannot corrupt the challenged accounts. Thus, if $\mathcal{A}$ distinguishes Game~3 from Game~2 with non-negligible probability, one can construct a standard IND-CPA adversary against the corresponding encryption primitive, using $\mathcal{A}$ as a subroutine and answering all oracle queries through $\mathcal{O}$ except the restricted corruption queries. Therefore,
	\begin{align*}
		\left|\Pr[E_3]-\Pr[E_2]\right|
		\leq
		\mathsf{negl}(\lambda).
	\end{align*}

	In Game~3, the challenge transaction is independent of the bit $b$ from the public adversarial view. Thus $\Pr[E_3]=1/2$. Combining the above game transitions yields
	\begin{align*}
		\left|\Pr[E_0]-\frac{1}{2}\right|
		\leq
		\mathsf{negl}(\lambda).
	\end{align*}
	Therefore, SilentLedger achieves anonymity.
\end{IEEEproof}

\begin{lemma} \label{lem:confidentiality}
	Assuming the IND-CPA security of \textsf{PKE} and \textsf{SKE}, the security of the keys derived from \textsf{NIKE} and \textsf{KDF}, and the zero-knowledge property of \textsf{SoK}, SilentLedger achieves confidentiality.
\end{lemma}

\begin{IEEEproof}[Proof]
	Let $E_0$ be the event that $\mathcal{A}$ wins the confidentiality experiment. In Game~0, $\mathcal{A}$ submits two admissible candidate amounts $\widehat{v}^0_1$ and $\widehat{v}^1_1$. For each $d\in\{0,1\}$, the complementary amount $v_1+v_2-\widehat{v}^d_1$ must lie in the valid range, so both branches satisfy the 2-2 balance condition.

	In Game~1, the challenger replaces the \textsf{SoK} proofs in the challenge transaction and in transactions returned by $\mathcal{O}_{tx}$ with simulated proofs. By the zero-knowledge property of \textsf{SoK},
	\begin{align*}
		\left|\Pr[E_1]-\Pr[E_0]\right|
		\leq
		\mathsf{negl}(\lambda).
	\end{align*}

	In Game~2, the challenger replaces the encrypted amount component corresponding to $\widehat{v}^b_1$ with an encryption of the alternative admissible amount. The public proof remains simulated, and the two branches have the same balance relation. If $\mathcal{A}$ distinguishes Game~2 from Game~1 with non-negligible probability, then one can build an IND-CPA adversary against \textsf{SKE}, because the changed value appears only inside the encrypted amount under a key unavailable to $\mathcal{A}$. Hence,
	\begin{align*}
		\left|\Pr[E_2]-\Pr[E_1]\right|
		\leq
		\mathsf{negl}(\lambda).
	\end{align*}

	In Game~3, the challenger replaces the ciphertexts for the recovery keys $CK_i$ and the output access keys $\widehat{c}_i$. Since the auditor's secret keys are not given to $\mathcal{A}$, and the challenge accounts cannot be corrupted, any non-negligible distinction between Game~3 and Game~2 would break the IND-CPA security of \textsf{PKE} or the security of the keys derived from \textsf{NIKE} and \textsf{KDF}. Therefore,
	\begin{align*}
		\left|\Pr[E_3]-\Pr[E_2]\right|
		\leq
		\mathsf{negl}(\lambda).
	\end{align*}

	In Game~3, all public values are independent of the challenge bit $b$. Hence $\Pr[E_3]=1/2$. Combining the above transitions gives
	\begin{align*}
		\left|\Pr[E_0]-\frac{1}{2}\right|
		\leq
		\mathsf{negl}(\lambda).
	\end{align*}
	Thus, SilentLedger achieves confidentiality.
\end{IEEEproof}

\begin{lemma} \label{lem:soundness}
	Assuming the soundness of \textsf{SoK} and the unforgeability of \textsf{RAC}, SilentLedger achieves soundness.
\end{lemma}

\begin{IEEEproof}[Proof]
	Let $E_0$ be the event that $\mathcal{A}$ wins the soundness experiment. That is, $\mathcal{A}$ outputs a transaction $tx=(\mathsf{stmt},m,\pi)$ such that $\mathsf{VerfTX}(pp,tx,X)=1$, but no witness $w$ satisfies $(\mathsf{stmt},w)\in\mathcal{L}_{tx}$.

	The public verification algorithm $\mathsf{VerfTX}$ accepts only if the \textsf{SoK} proof verifies and the adapted signatures $\widehat{\sigma}'_i$ for the output anonymous addresses $\widehat{Q}_i$ pass public verification. Therefore, if $tx$ is accepted while no valid witness exists for $\mathsf{stmt}$, then either the \textsf{SoK} proof is accepted for a false statement, or an adapted certificate verifies without a valid certified origin. The first case contradicts the soundness of \textsf{SoK}, and the second case contradicts the unforgeability of \textsf{RAC}. Hence,
	\begin{align*}
		\Pr[E_0]\leq \mathsf{negl}(\lambda).
	\end{align*}
	SilentLedger therefore achieves soundness.
\end{IEEEproof}
	By Lemmas~\ref{lem:authenticity}, \ref{lem:anonymity}, \ref{lem:confidentiality}, and \ref{lem:soundness}, SilentLedger satisfies all four security properties. This completes the proof of Theorem~\ref{security}.
\end{IEEEproof}

\section{An Efficient Instantiation} \label{Instantiation}

This section presents efficient instantiations of the cryptographic primitives, the SoK relations, and the system workflow under the 2-2 transaction model.
Let $(\mathbb{G}_1,\mathbb{G}_2,\mathbb{G}_T,e)$ be bilinear groups of prime order $q$, with $e:\mathbb{G}_1\times\mathbb{G}_2\to\mathbb{G}_T$ and generators $G_1\in\mathbb{G}_1$ and $G_2\in\mathbb{G}_2$.

\subsection{Instantiating Cryptographic Primitives}
To complement the RAC scheme, SilentLedger instantiates \textsf{PKE} with elliptic-curve ElGamal and uses the following efficient primitives.

\begin{itemize}
\item $pp \gets \mathsf{PKESetup}(\lambda)$: Choose an elliptic-curve group $\mathbb{G}_1$ of prime order $q$ with generator $G_1$, and output $pp=(\mathbb{G}_1,q,G_1)$.
\item $(sk,pk)\gets\mathsf{PKEKGen}(pp)$: Sample $sk\gets\mathbb{Z}_q$ and set $pk=skG_1$.
\item $CK\gets\mathsf{PKEEnc}(pp,m,pk)$: For a plaintext $m\in\mathbb{G}_1$, sample the encryption randomness $\gamma\gets\mathbb{Z}_q$ and output the elliptic-curve ElGamal ciphertext $CK=(C,D)=(\gamma G_1,m+\gamma pk)$.
\item $m\gets\mathsf{PKEDec}(CK,sk)$: For $CK=(C,D)$, recover $m=D-skC$.
\end{itemize}
In SilentLedger, the encrypted plaintext is the recovery key $\widehat{K}_i$, and the auditor's public key is $T$. Thus, \textsf{PKE} generates the ciphertext $CK_i=(C_i,D_i)=(\gamma_iG_1,\widehat{K}_i+\gamma_iT)$.
We then instantiate \textsf{SKE} as an encryption scheme for transaction data:
\begin{itemize}
\item $c \gets \mathsf{SKEnc}(pp, m, xk)$: Encrypts message $m$ with symmetric key $xk$, outputs $c = m + xk$.
\item $m \gets \mathsf{SKDec}(c, xk)$: Recovers $m = c - xk$.

In SilentLedger, \textsf{SKEnc} hides the account address as $\widehat{Q}_i = \widehat{S}_i + \widehat{K}_i$ and the encoded amount as $\widehat{cm}_i = \widehat{vx}_i + \widehat{tc}_i$.
\end{itemize}

For \textsf{NIKE}, we use Diffie--Hellman non-interactive key derivation over $\mathbb{G}_1$. Let a hash function $\mathcal{H}$ map $\mathbb{G}_1$ elements to $\mathbb{Z}_q$.  In SilentLedger,  the payer derives the access key as $\widehat{c}_i = \mathsf{NIKE}(\widehat{V}_i, r_i)=\mathcal{H}(\widehat{V}_i \cdot r_i)=\mathcal{H}(\widehat{vk}_i \cdot R_i)$. The symmetric key is $\widehat{tc}_i = \mathsf{NIKE}(T, \widehat{c}_i)=\mathcal{H}(T \cdot \widehat{c}_i)=\mathcal{H}(mk \cdot \widehat{K}_i)$. Thus, the auditor derives the recovery key from $mk$ and the posted $\widehat{K}_i$, and payees can access keys from their secret key $\widehat{v}_i$ and the posted $R_i$.

\textsf{KDF} is instantiated as $\widehat{K}_i = \widehat{c}_i G_1$. Hence, the anonymous address is $\widehat{Q}_i = \widehat{S}_i + \widehat{K}_i = \widehat{S}_i + \widehat{c}_i G_1$, which is consistent with the RAC adaptation relation $\mathsf{RAC.Rndmz}(\widehat{S}_i, \widehat{c}_i)$.

The reversible function \textsf{RF} encodes $\widehat{v}_i$ as $\widehat{vx}_i=\widehat{v}_i G_1$. Since $\widehat{v}_i$ lies in a bounded range, it can be recovered from $\widehat{vx}_i$ by a bounded discrete-log search.

\subsection{Instantiating SoK}

Under the 2-2 transaction model, SilentLedger uses two \textsf{SoK} relations: $\mathcal{L}_{tx_1}$ for transaction correctness, audit binding, and certificate consistency, and $\mathcal{L}_{tx_2}$ for output amount range validity.

\textbf{Design of \textsf{SoK} for $\mathcal{L}_{tx_1}$.}
To satisfy the verification requirements of the above construction, it is necessary to prevent transaction parties from colluding to bypass auditing. In particular, public verification of an adapted certificate only authenticates the posted anonymous address $\widehat{Q}_i$. It does not by itself prove that the access key $\widehat{c}_i$ used in the ciphertext for the recovery key is exactly the randomization factor used to derive $\widehat{Q}_i$ from a certified account address. Otherwise, a malicious payer could post a valid adapted certificate for $\widehat{Q}_i$ while choosing an inconsistent $\widehat{c}_i$, causing the auditor to recover a fabricated account address.

Therefore, $\mathcal{L}_{tx_1}$ must bind the same value $\widehat{c}_i$ to the encrypted amount $\widehat{cm}_i$, the ciphertext for the recovery key $CK_i$, and the RAC certificate adaptation $(\widehat{Q}_i, \widehat{\sigma}'_i)$. Let
\[
\widehat{\sigma}_i=(Z_i,U_i,\widetilde{U}_i,Y_i),
\qquad
\widehat{\sigma}'_i=(Z_i',U_i',\widetilde{U}_i',Y_i').
\]
The relation for adapting certificates is written as
\[
\mathsf{AdaptRel}(X,\widehat{Q}_i,\widehat{\sigma}'_i,\widehat{c}_i,
\widehat{S}_i,\widehat{\sigma}_i,s_i')=1,
\]
which means that $\widehat{\sigma}_i$ is a valid RAC certificate for the hidden account address $\widehat{S}_i$, the anonymous address satisfies
$
	\widehat{Q}_i=\widehat{S}_i+\widehat{c}_iG_1,
$
and, writing
$
\widehat{\sigma}_i=(Z_i,U_i,\widetilde{U}_i,Y_i),
\widehat{\sigma}'_i=(Z_i',U_i',\widetilde{U}_i',Y_i'),
$
the adapted certificate satisfies
\[
s_i'Z_i'=Z_i+\widehat{c}_iY_i,\quad
U_i'=s_i'U_i,\quad
\widetilde{U}_i'=s_i'\widetilde{U}_i,\quad
Y_i=s_i'Y_i'.
\]
The public verifier additionally checks $\mathsf{Verify}(X,\widehat{Q}_i,\widehat{\sigma}'_i)=1$.

Accordingly, $\mathcal{L}_{tx_1}$ is defined as follows:
\begin{align*}
\resizebox{\columnwidth}{!}{%
$\mathcal{L}_{tx_1}=\mathsf{SoK} \left\{
\begin {array}{l}
(\mathsf{stmt}_1,w_1): \\
	cm_1 ={v_1}G_1 + {c_1}T \land cm_2 = {v_2}G_1 + {c_2}T \land \\
	\widehat{cm}_1 = {\widehat{v}_1}G_1 + {\widehat{c}_1}T \land
	\widehat{cm}_2 = {\widehat{v}_2}G_1 + {\widehat{c}_2}T \land \\
	Q_1 = (s_1+c_1)G_1 \land Q_2 = (s_2+c_2)G_1 \land \\
	\widehat{Q}_1=\widehat{S}_1+\widehat{c}_1G_1 \land
	\widehat{Q}_2=\widehat{S}_2+\widehat{c}_2G_1 \land\\
\mathsf{AdaptRel}(X,\widehat{Q}_1,\widehat{\sigma}'_1,\widehat{c}_1,
\widehat{S}_1,\widehat{\sigma}_1,s_1')=1 \land\\
\mathsf{AdaptRel}(X,\widehat{Q}_2,\widehat{\sigma}'_2,\widehat{c}_2,
\widehat{S}_2,\widehat{\sigma}_2,s_2')=1 \land\\
	C_1=\gamma_1G_1 \land C_2=\gamma_2G_1 \land\\
	D_1=\widehat{c}_1G_1+\gamma_1T \land
	D_2=\widehat{c}_2G_1+\gamma_2T \land\\
(cm_1+cm_2)-(\widehat{cm}_1+\widehat{cm}_2)
=(c_1+c_2-\widehat{c}_1-\widehat{c}_2)T
\end{array}
\right\}(m).$%
}
\end{align*}
where
$
\mathsf{stmt}_1=(cm_1,cm_2,\widehat{cm}_1,\widehat{cm}_2,Q_1,Q_2,
\widehat{Q}_1,\widehat{Q}_2,C_1,C_2,\\
D_1,D_2,G_1,T,X,\widehat{\sigma}'_1,\widehat{\sigma}'_2)
$
and
$
w_1=(v_1,v_2,c_1,c_2,s_1,s_2,\\
\widehat{v}_1,\widehat{v}_2,\widehat{c}_1,\widehat{c}_2,
\widehat{S}_1,\widehat{S}_2,\gamma_1,\gamma_2,\widehat{\sigma}_1,\widehat{\sigma}_2,s_1',s_2').
$
The adapted RAC signatures $\widehat{\sigma}'_1$ and $\widehat{\sigma}'_2$ are publicly verified by $\mathsf{Verify}$, while the above relation binds their adaptation randomness to the same access keys $\widehat{c}_1$ and $\widehat{c}_2$ used in the ciphertexts for the recovery keys. This binding is the core reason why a malicious payer cannot post a valid adapted RAC signature while encrypting an inconsistent access key for the auditor.

To instantiate this relation, the proof $\pi_1$ is obtained by applying the Fiat--Shamir transform to a composition of two subproofs. The first subproof, denoted by $\pi_{\mathsf{lin}}$, proves the linear relations among commitments, anonymous addresses, ciphertexts, and the balance equation. The second subproof, denoted by $\pi_{\mathsf{rac}}$, proves the RAC certificate without revealing the certified account address.

For $\pi_{\mathsf{lin}}$, for each $i\in\{1,2\}$, the prover samples
$
r_{v_i},r_{c_i},r_{s_i},r_{\widehat{v}_i},r_{\widehat{c}_i},r_{\gamma_i}
\gets \mathbb{Z}_q$,
$
R_{\widehat{S}_i}\gets\mathbb{G}_1.
$
It computes the commitments
\begin{align*}
T_{cm_i} &= r_{v_i}G_1+r_{c_i}T, \qquad&
T_{\widehat{cm}_i} &= r_{\widehat{v}_i}G_1+r_{\widehat{c}_i}T,\\
T_{Q_i} &= (r_{s_i}+r_{c_i})G_1,\qquad&
T_{\widehat{Q}_i} &= R_{\widehat{S}_i}+r_{\widehat{c}_i}G_1,\\
T_{C_i} &= r_{\gamma_i}G_1,\qquad&
T_{D_i} &= r_{\widehat{c}_i}G_1+r_{\gamma_i}T.
\end{align*}
For the balance equation, it computes
\[
T_{\mathsf{bal}}
=
(r_{c_1}+r_{c_2}-r_{\widehat{c}_1}-r_{\widehat{c}_2})T.
\]
The challenge is computed as
\begingroup
\small
\[
e_1=\mathcal{H}(\mathsf{stmt}_1,m,\{T_{cm_i},T_{\widehat{cm}_i},T_{Q_i},
T_{\widehat{Q}_i},T_{C_i},T_{D_i}\}_{i=1}^{2},
T_{\mathsf{bal}},\pi_{\mathsf{rac}}).
\]
\endgroup
The prover returns the responses
\begin{align*}
z_{v_i} &= r_{v_i}-e_1v_i, &
z_{c_i} &= r_{c_i}-e_1c_i,\\
z_{s_i} &= r_{s_i}-e_1s_i, &
z_{\widehat{v}_i} &= r_{\widehat{v}_i}-e_1\widehat{v}_i,\\
z_{\widehat{c}_i} &= r_{\widehat{c}_i}-e_1\widehat{c}_i, &
z_{\gamma_i} &= r_{\gamma_i}-e_1\gamma_i,\\
Z_{\widehat{S}_i} &= R_{\widehat{S}_i}-e_1\widehat{S}_i.
\end{align*}
The verifier recomputes
\begingroup
\small
\begin{align*}
T'_{cm_i} &= z_{v_i}G_1+z_{c_i}T+e_1cm_i,
&T'_{\widehat{cm}_i} &= z_{\widehat{v}_i}G_1+z_{\widehat{c}_i}T+e_1\widehat{cm}_i,\\
T'_{Q_i} &= (z_{s_i}+z_{c_i})G_1+e_1Q_i,
&T'_{\widehat{Q}_i} &= Z_{\widehat{S}_i}+z_{\widehat{c}_i}G_1+e_1\widehat{Q}_i,\\
T'_{C_i} &= z_{\gamma_i}G_1+e_1C_i,
&T'_{D_i} &= z_{\widehat{c}_i}G_1+z_{\gamma_i}T+e_1D_i,
\end{align*}
\endgroup
\[
T'_{\mathsf{bal}}
=
(z_{c_1}+z_{c_2}-z_{\widehat{c}_1}-z_{\widehat{c}_2})T
+
e_1\big((cm_1+cm_2)-(\widehat{cm}_1+\widehat{cm}_2)\big).
\]
The verifier accepts $\pi_{\mathsf{lin}}$ if the recomputed challenge equals $e_1$.

The subproof $\pi_{\mathsf{rac}}$ proves that, for each output, there exists a hidden account address $\widehat{S}_i$, a valid RAC certificate $\widehat{\sigma}_i$, and an adaptation scalar $s_i'$ such that $\mathsf{AdaptRel}(X,\widehat{Q}_i,\widehat{\sigma}'_i,\widehat{c}_i,\widehat{S}_i,\widehat{\sigma}_i,s_i')=1$. It is composed with $\pi_{\mathsf{lin}}$ under the same Fiat--Shamir challenge, and \textsf{VerfTX} additionally checks $\mathsf{Verify}(X,\widehat{Q}_i,\widehat{\sigma}'_i)=1$.

\textbf{Design of \textsf{SoK} for $\mathcal{L}_{tx_2}$}:
$\mathcal{L}_{tx_2}$ proves that the output amounts lie in the valid range. Formally,
\[
\mathcal{L}_{tx_2}
=
\mathsf{SoK}\{(\mathsf{stmt}_2,w_2):
\widehat{v}_1,\widehat{v}_2\in[0,2^{n-1})\}(m),
\]
where $\mathsf{stmt}_2=(T,n,\widehat{cm}_1,\widehat{cm}_2)$ and
$w_2=(\widehat{v}_1,\widehat{v}_2,\widehat{c}_1,\widehat{c}_2)$.

We instantiate this relation using Bulletproofs++~\cite{Bulletproofs++}. The prover generates an aggregated range proof for the two output commitments:
\[
\pi_2 \leftarrow
\mathsf{BP.Prove}(\{\widehat{cm}_i\}_{i=1}^{2},T,\vec{g},\vec{h},
\{(\widehat{v}_i,\widehat{c}_i)\}_{i=1}^{2}).
\]
The verifier checks
$
b \leftarrow
\mathsf{BP.Verf}(\{\widehat{cm}_i\}_{i=1}^{2},T,\vec{g},\vec{h},\pi_2),
$
and accepts if $b=1$. Both $\pi_1$ and $\pi_2$ are made non-interactive using the Fiat--Shamir transform.

\begin{definition}
For $\mathcal{L}_{tx}=\mathcal{L}_{tx_1}\land\mathcal{L}_{tx_2}$, the instantiated transaction \textsf{SoK} is secure if it satisfies \textit{completeness}, \textit{knowledge soundness}, \textit{zero-knowledge}, and \textit{message unforgeability}, where no $\mathcal{PPT}$ adversary can output an accepting proof for a fresh message-statement pair without a valid witness.
\end{definition}

\begin{theorem}
\label{thm:instantiated-sok-security}
Assume that $\pi_{\mathsf{lin}}$ is obtained from a $\Sigma$-protocol satisfying special soundness and honest-verifier zero-knowledge~\cite{Proof_systems_for_general_statements_about_discrete_logarithms}, the Fiat--Shamir transform is secure in the random oracle model~\cite{How_to_prove_yourself}, Bulletproofs++ is knowledge-sound and zero-knowledge~\cite{Bulletproofs++}, and \textsf{RAC} is unforgeable as proved in Section~\ref{sec:rac_security_analysis}. The instantiated transaction \textsf{SoK} satisfies completeness, knowledge soundness, zero-knowledge, and message unforgeability.
\end{theorem}

\begin{IEEEproof}[Proof Sketch]
Completeness follows from the correctness of the linear relations in $\mathcal{L}_{tx_1}$, the RAC adaptation relation, and Bulletproofs++ for $\mathcal{L}_{tx_2}$.

For knowledge soundness, suppose an adversary outputs an accepting proof $\pi=(\pi_1,\pi_2)$. By applying the forking lemma to the Fiat--Shamir transform~\cite{Security_Arguments_for_Digital_Signatures_and_Blind_Signatures}, two accepting transcripts of $\pi_{\mathsf{lin}}$ with the same commitments and different challenges allow extraction of
$
(v_i,c_i,s_i,\widehat{v}_i,\widehat{c}_i,\widehat{S}_i,\gamma_i),
$
which satisfies the commitment, anonymous address, ciphertext for the recovery key, and balance equations in $\mathcal{L}_{tx_1}$. The soundness of $\pi_{\mathsf{rac}}$ binds the adapted certificate to the same $\widehat{c}_i$, while the unforgeability of \textsf{RAC} prevents certificates for uncertified account addresses. The knowledge soundness of Bulletproofs++ extracts valid openings $(\widehat{v}_i,\widehat{c}_i)$ and guarantees $\widehat{v}_i\in[0,2^{n-1})$. Thus, any adversary breaking knowledge soundness can be reduced to breaking the special soundness of the linear $\Sigma$-protocol, the soundness of $\pi_{\mathsf{rac}}$, the unforgeability of \textsf{RAC}, or the knowledge soundness of Bulletproofs++.

Zero-knowledge follows from the simulators of the underlying $\Sigma$-protocols and Bulletproofs++~\cite{Bulletproofs++}. Under the random oracle model, the Fiat--Shamir challenges can be simulated in the standard way. Message unforgeability follows because $m$ is included in the Fiat--Shamir challenge: a proof cannot be transferred to a fresh pair of message and statement unless the adversary finds a collision in the random oracle or produces an accepting proof without an extractable witness.
\end{IEEEproof}

\subsection{Instantiating SilentLedger}

We now describe the instantiated SilentLedger workflow under the 2-2 transaction model.

\begin{figure}[!htb]
    \centering
    \includegraphics[width=\columnwidth]{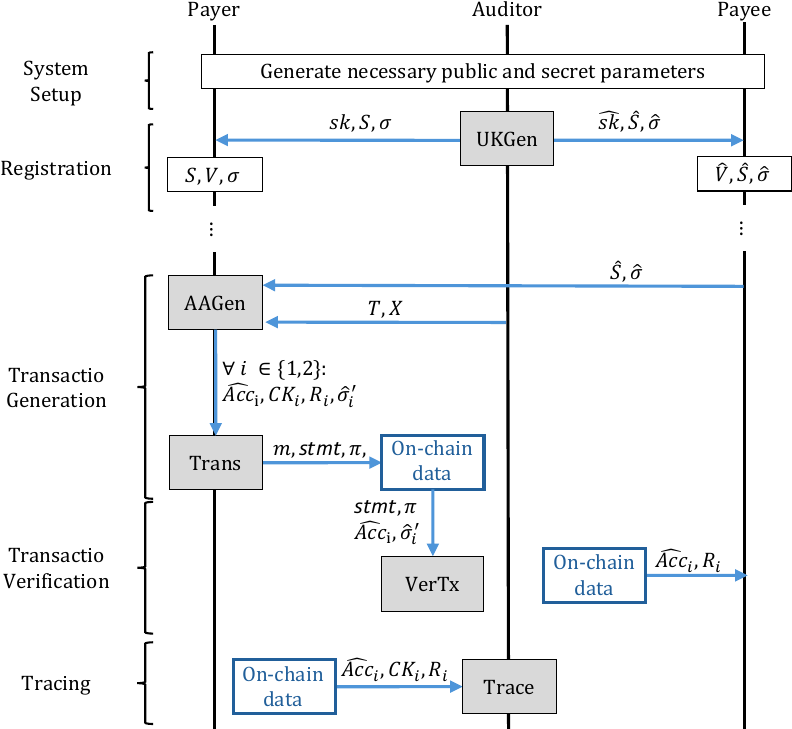}
    \caption{The workflow of SilentLedger.}
    \label{fig:overview_silentledger}
\end{figure}

\textbf{System assumptions.}
SilentLedger operates in four phases: (i) registration phase, (ii) transaction generation phase, (iii) transaction verification phase, and (iv) tracing phase. Fig.~\ref{fig:overview_silentledger} summarizes the local keys and interactions among the payer, payee, and auditor. The auditor acts as a trusted third party for certification and tracing, and maintains an online public directory for registered account addresses and certificates. This directory is a service for system management and does not require auditor participation in transaction generation or verification.

\textbf{SilentLedger Design Details.} We summarize the workflow of SilentLedger across the four phases.

\underline{Registration phase:} 
Each user registers an account address with the auditor, who issues a RAC certificate $(S,\sigma)$ binding the user's identity to $S$. The user also generates a viewing key pair $(vk,V)$ and publishes $(S,V,\sigma)$ as the public registration record for receiving future transactions.

\underline{Transaction generation phase:} Payer creates a transaction $tx$ distributing $\widehat{v}_1$ and $\widehat{v}_2$ to two outputs. For each output, the payer generates an anonymous account $\widehat{Acc}_i = (\widehat{Q}_i, \widehat{cm}_i)$,  corresponding adapted signature $\widehat{\sigma}'_i$ , a ciphertext for the recovery key $CK_i$, and a ephemeral public key $R_i$. The, the Payer generates \textsf{SoK} proofs $\pi$ of transaction message for transaction verification. Finally, the transaction $tx=(\mathsf{stmt},m,\pi)$ is posted.

To create a 2-2 transaction, the payer spends two input accounts and creates two output accounts carrying amounts $\widehat{v}_1$ and $\widehat{v}_2$. For each output, the payer uses the payee's public registration record to generate an anonymous account $\widehat{Acc}_i=(\widehat{Q}_i,\widehat{cm}_i)$, an adapted RAC signature $\widehat{\sigma}'_i$, an recovery ciphertext $CK_i$, and an ephemeral public key $R_i$. The payer then posts $tx=(\mathsf{stmt},m,\pi)$, where $\pi$ proves transaction correctness and The payer then posts $tx=(\mathsf{stmt},m,\pi)$, where $\pi$ proves transaction correctness and auditable.

\underline{Transaction verification phase:} Validation nodes verify the posted transaction using only public data. They check the adapted RAC signatures and the \textsf{SoK} proof, and accept the transaction only if all verification checks pass.

\underline{Tracing phase:} When authorized tracing is required, the auditor uses its tracing secret key and the on-chain data $(\widehat{Acc}_i,CK_i,R_i)$ to recover the payee's registered account address and transaction amount. 

\section{Evaluation and Comparison}

This section evaluates SilentLedger and compares it with related auditable payment systems, including Platypus~\cite{Platypus}, PEReDi~\cite{Peredi}, and PGC~\cite{PGC}.
Experiments were conducted on Windows~11 with an Intel Core i7-10500 @ 2.90\,GHz and 16\,GB RAM. We implemented cryptographic operations using mcl~\cite{mcl} over BLS12-381 with asymmetric pairing $e:\mathbb{G}_1\times\mathbb{G}_2\rightarrow\mathbb{G}_T$. The implementation targets the 128-bit classical security level. We select BLS12-381 because it supports efficient Type-3 pairings and prime-order subgroups~~\cite{BLS12_381,mcl}. The descriptions of its elements and relevant bit lengths are shown in Table~\ref{tab:parameters}.

\textbf{Implementation.}
SilentLedger was implemented in C++, with Bulletproofs++~\cite{Bulletproofs++} used for $\mathcal{L}_{tx_2}$. Table~\ref{tab:parameters} summarizes the parameters used in the time and size analysis. The theoretical costs count dominant elliptic-curve and pairing operations and omit lightweight hashing and modular arithmetic. Consequently, measured runtimes are slightly higher than theoretical estimates.

\begin{table}[!b]
	\centering
	\caption{Parameters for performance analysis.}
	\label{tab:parameters}
	\normalsize
	\setlength{\tabcolsep}{2.5pt}
	\renewcommand{\arraystretch}{1.06}
	\begin{tabularx}{\columnwidth}{@{}>{\centering\arraybackslash}c>{\raggedright\arraybackslash}X>{\centering\arraybackslash}c@{}}
		\toprule
		\textbf{Parameter} & \textbf{Description} & \textbf{Size} \\
		\midrule
		\multicolumn{3}{@{}l}{\textit{Size parameters}}\\
	$|p_{\mathrm{BLS}}|$ & Base-field characteristic of BLS12-381 & 381 bits \\
	$|q|$ & Subgroup order of BLS12-381 & 255 bits \\
	$|\mathbb{G}_1|$ & Element length in $\mathbb{G}_1$ & 762 bits \\
	$|\mathbb{G}_2|$ & Element length in $\mathbb{G}_2$ & 1524 bits \\
	$|\mathbb{G}_T|$ & Element length in $\mathbb{G}_T$ & 4572 bits\\
	$|v|$ & Length of the transaction amount & 32 bits \\
		\midrule
		\multicolumn{3}{@{}l}{\textit{Time parameters}}\\
		$T_{G_1}$ & Time of a point multiplication in $\mathbb{G}_1$ & -- \\
		$T_{G_2}$ & Time of a point multiplication in $\mathbb{G}_2$ & -- \\
		$T_{G_T}$ & Time of a multiplication in $\mathbb{G}_T$ & -- \\
		$T_b$ & Time of a bilinear pairing operation & -- \\
		$T_{bsgs}$ & Time of baby-step giant-step algorithm & -- \\
		\bottomrule
	\end{tabularx}
\end{table}

\begin{table}[!b]
	\centering
	\caption{Algorithm runtime and size costs of SilentLedger.}
	\label{tab:runtime_size_costs}
	\normalsize
	\setlength{\tabcolsep}{2pt}
	\renewcommand{\arraystretch}{1.06}
\begin{tabular}{@{}ccc@{}}
    \toprule
    \textbf{Item} & \textbf{Theory} & \textbf{Practice} \\
    \midrule
    \multicolumn{3}{@{}l}{\textit{Algorithm runtime}}\\
    \textsf{Setup} & -- & 5.49 ms \\
    \textsf{MKGen} & $T_{G_1}+T_{G_2}$ & 1.83 ms \\
    \textsf{UKGen} & $5T_{G_1}+T_{G_2}$ & 1.04 ms \\
    \textsf{AAGen} & $14T_{G_1}$ & 1.88 ms \\
    \textsf{Trans} & $(4n+33)T_{G_1}+2T_{G_2}$ & 9.75 ms \\
    \textsf{VerfTX} & $(6n+38)T_{G_1}+14T_b+2T_{G_T}$ & 22.43 ms \\
    \textsf{Trace} & $T_{G_1}+T_{\mathrm{bsgs}}$ & 2.03 s \\
    \midrule
    \multicolumn{3}{@{}l}{\textit{Data size}}\\
    Account address & $2|\mathbb{G}_1|$ & 288 B \\
    Transaction & $(27+2\log_2 n)|\mathbb{G}_1|+|\mathbb{G}_2|$ & 4896 B \\
    \bottomrule
\end{tabular}
\end{table}

\begin{figure}[!b]
	\centering
	\includegraphics[width=0.46\textwidth]{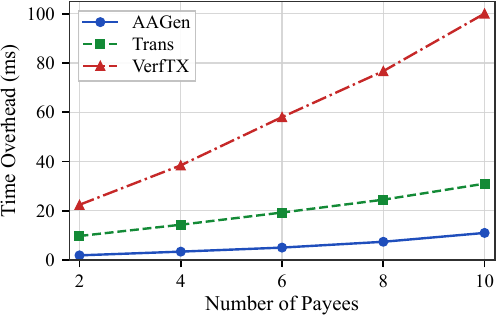}
	\caption{Time overhead of \textsf{AAGen}, \textsf{Trans}, and \textsf{VerfTX}.}
	\label{fig:time_cost}
\end{figure}

\textbf{Computational Overhead.} Each algorithm was run 10,000 times under the 2-2 transaction model with all amounts set to 1 million. As shown in Table~\ref{tab:runtime_size_costs}, \textsf{Trans} and \textsf{VerfTX} take 9.75 ms and 22.43 ms, respectively. \textsf{Trace} takes 2.03 s due to bounded discrete-log recovery. Fig.~\ref{fig:time_cost} further shows that \textsf{AAGen} and \textsf{Trans} remain below 40 ms as the number of payees grows. For \textsf{VerfTX}, validation nodes usually have sufficient computing and storage resources, so the verification overhead remains within acceptable limits. These costs are small compared with typical blockchain confirmation delays (e.g., 12 seconds for Ethereum).

\textbf{Comparison.}
We compare SilentLedger with Platypus, PEReDi, and PGC using three metrics that directly affect blockchain practicality: Transaction Generation Time, Transaction Validation Time, and Transaction Size. Following prior work~\cite{Platypus, Peredi, PGC}, we use the Platypus test data under the holding-limit and receiving-limit settings. These settings closely match our protocol's operational requirements. For PEReDi, we configured the environment variables $n_v$, $n_s$, and $n_b$ to 32, following the recommended settings from~\cite{Peredi}.

\begin{figure}[!t]
	\includegraphics[width=\columnwidth]{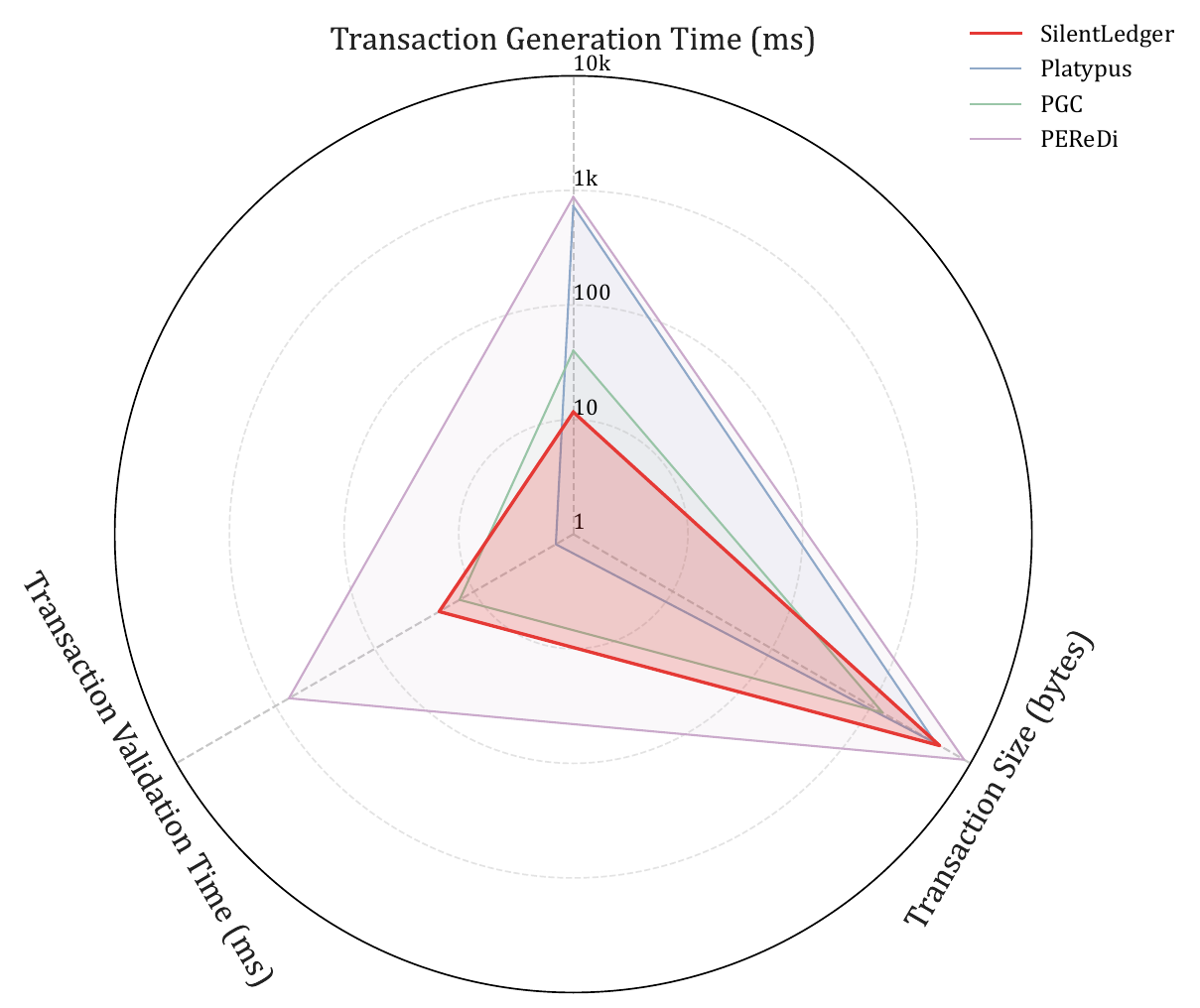}
	\caption{Performance comparison of SilentLedger with Platypus, PGC, and PEReDi using a logarithmic radial scale.}
	\label{fig:radar_chart}
\end{figure}

Since the metrics use different units and ranges, Fig.~\ref{fig:radar_chart} uses a logarithmic radial scale. The results show that SilentLedger achieves the lowest transaction generation time, 11.63 ms, compared with 40 ms for PGC, 730 ms for Platypus, and 877.51 ms for PEReDi. PGC and Platypus have lower validation times (14 ms and 1.5 ms versus 22.43 ms for SilentLedger) and smaller transaction sizes (1310 bytes and 4488 bytes versus 4896 bytes for SilentLedger), but SilentLedger remains practical while providing stronger auditing functionality. More importantly, compared with PEReDi, SilentLedger reduces transaction generation time by 99.87\% (from 877.51 ms to 11.63 ms), validation time by 96.94\% (from 731.9 ms to 22.43 ms), and transaction size by 56.7\% (from 8632 bytes to 4896 bytes).

Furthermore, unlike Platypus and PEReDi, which require auditor participation during transaction execution, SilentLedger supports independent transaction processing. Compared with PGC, which supports confidential payments and non-interactive amount auditing, SilentLedger additionally provides identity anonymity and authorized identity recovery with comparable validation overhead. The modest increase in transaction size relative to some competing solutions is attributable to our novel identity management mechanism, which ensures robust authentication. This trade-off is well justified by the stronger identity-privacy protection and additional auditing capabilities provided by our system.

\section{Conclusion}

We presented SilentLedger, a privacy-preserving auditing system that removes the interaction dependencies typically imposed by audit mechanisms. By decoupling transaction execution from auditors and auditing from users, SilentLedger achieves complete non-interactivity. We introduce RAC  to derive publicly verifiable certificates for anonymous account addresses without revealing registered identities, while the transaction structure enables payees and auditors to recover the data they need from their own keys and on-chain data.
We formally analyze its security and give an implementation to show practical performance compared with representative auditable payment systems.
\bibliographystyle{IEEEtran}
\bibliography{SilentLedger}

\begin{thebibliography}{10}
\providecommand{\url}[1]{#1}
\csname url@samestyle\endcsname
\providecommand{\newblock}{\relax}
\providecommand{\bibinfo}[2]{#2}
\providecommand{\BIBentrySTDinterwordspacing}{\spaceskip=0pt\relax}
\providecommand{\BIBentryALTinterwordstretchfactor}{4}
\providecommand{\BIBentryALTinterwordspacing}{\spaceskip=\fontdimen2\font plus
\BIBentryALTinterwordstretchfactor\fontdimen3\font minus \fontdimen4\font\relax}
\providecommand{\BIBforeignlanguage}[2]{{%
\expandafter\ifx\csname l@#1\endcsname\relax
\typeout{** WARNING: IEEEtran.bst: No hyphenation pattern has been}%
\typeout{** loaded for the language `#1'. Using the pattern for}%
\typeout{** the default language instead.}%
\else
\language=\csname l@#1\endcsname
\fi
#2}}
\providecommand{\BIBdecl}{\relax}
\BIBdecl

\bibitem{The_truth_about_blockchain}
M.~Iansiti, K.~R. Lakhani \emph{et~al.}, ``The truth about blockchain,'' \emph{Harvard business review}, vol.~95, no.~1, pp. 118--127, 2017.

\bibitem{DCAP}
C.~Lin, D.~He, X.~Huang, M.~K. Khan, and K.-K.~R. Choo, ``Dcap: A secure and efficient decentralized conditional anonymous payment system based on blockchain,'' \emph{IEEE Transactions on Information Forensics and Security}, vol.~15, pp. 2440--2452, 2020.

\bibitem{Ppchain}
C.~Lin, D.~He, X.~Huang, X.~Xie, and K.-K.~R. Choo, ``Ppchain: A privacy-preserving permissioned blockchain architecture for cryptocurrency and other regulated applications,'' \emph{IEEE Systems Journal}, vol.~15, no.~3, pp. 4367--4378, 2021.

\bibitem{Bulletproofs}
B.~B{\"u}nz, J.~Bootle, D.~Boneh, A.~Poelstra, P.~Wuille, and G.~Maxwell, ``Bulletproofs: Short proofs for confidential transactions and more,'' in \emph{2018 IEEE symposium on security and privacy (SP)}.\hskip 1em plus 0.5em minus 0.4em\relax IEEE, 2018, pp. 315--334.

\bibitem{CryptoNote2}
N.~Van~Saberhagen, ``Cryptonote v 2.0,'' 2013.

\bibitem{Zerocash}
E.~B. Sasson, A.~Chiesa, C.~Garman, M.~Green, I.~Miers, E.~Tromer, and M.~Virza, ``Zerocash: Decentralized anonymous payments from bitcoin,'' in \emph{2014 IEEE symposium on security and privacy}.\hskip 1em plus 0.5em minus 0.4em\relax IEEE, 2014, pp. 459--474.

\bibitem{Ringct_2.0}
S.-F. Sun, M.~H. Au, J.~K. Liu, and T.~H. Yuen, ``Ringct 2.0: A compact accumulator-based (linkable ring signature) protocol for blockchain cryptocurrency monero,'' in \emph{Computer Security--ESORICS 2017: 22nd European Symposium on Research in Computer Security, Oslo, Norway, September 11-15, 2017, Proceedings, Part II 22}.\hskip 1em plus 0.5em minus 0.4em\relax Springer, 2017, pp. 456--474.

\bibitem{Aggregate_cash_systems}
G.~Fuchsbauer, M.~Orr{\`u}, and Y.~Seurin, ``Aggregate cash systems: A cryptographic investigation of mimblewimble,'' in \emph{Advances in Cryptology--EUROCRYPT 2019: 38th Annual International Conference on the Theory and Applications of Cryptographic Techniques, Darmstadt, Germany, May 19--23, 2019, Proceedings, Part I 38}.\hskip 1em plus 0.5em minus 0.4em\relax Springer, 2019, pp. 657--689.

\bibitem{Quisquis}
P.~Fauzi, S.~Meiklejohn, R.~Mercer, and C.~Orlandi, ``Quisquis: A new design for anonymous cryptocurrencies,'' in \emph{Advances in Cryptology--ASIACRYPT 2019: 25th International Conference on the Theory and Application of Cryptology and Information Security, Kobe, Japan, December 8--12, 2019, Proceedings, Part I 25}.\hskip 1em plus 0.5em minus 0.4em\relax Springer, 2019, pp. 649--678.

\bibitem{Non-interactive_mimblewimble}
G.~Fuchsbauer and M.~Orr{\`u}, ``Non-interactive mimblewimble transactions, revisited,'' in \emph{International Conference on the Theory and Application of Cryptology and Information Security}.\hskip 1em plus 0.5em minus 0.4em\relax Springer, 2022, pp. 713--744.

\bibitem{pedersen}
T.~P. Pedersen, ``Non-interactive and information-theoretic secure verifiable secret sharing,'' in \emph{Annual international cryptology conference}.\hskip 1em plus 0.5em minus 0.4em\relax Springer, 1991, pp. 129--140.

\bibitem{Blockchain_Regulatory_Certainty_Act}
U.~S. Congress, ``Blockchain regulatory certainty act,'' \url{https://www.congress.gov/bill/117th-congress/house-bill/5045}, 2023.

\bibitem{Traceable_monero}
Y.~Li, G.~Yang, W.~Susilo, Y.~Yu, M.~H. Au, and D.~Liu, ``Traceable monero: Anonymous cryptocurrency with enhanced accountability,'' \emph{IEEE Transactions on Dependable and Secure Computing}, vol.~18, no.~2, pp. 679--691, 2021.

\bibitem{Platypus}
K.~W{\"u}st, K.~Kostiainen, N.~Delius, and S.~Capkun, ``Platypus: A central bank digital currency with unlinkable transactions and privacy-preserving regulation,'' in \emph{Proceedings of the 2022 ACM SIGSAC Conference on Computer and Communications Security}, 2022, pp. 2947--2960.

\bibitem{Aca}
C.~Lin, X.~Huang, J.~Ning, and D.~He, ``Aca: Anonymous, confidential and auditable transaction systems for blockchain,'' \emph{IEEE Transactions on Dependable and Secure Computing}, vol.~20, no.~6, pp. 4536--4550, 2023.

\bibitem{Peredi}
A.~Kiayias, M.~Kohlweiss, and A.~Sarencheh, ``Peredi: Privacy-enhanced, regulated and distributed central bank digital currencies,'' in \emph{Proceedings of the 2022 ACM SIGSAC Conference on Computer and Communications Security}, 2022, pp. 1739--1752.

\bibitem{zkCross}
Y.~Guo, M.~Xu, X.~Cheng, D.~Yu, W.~Qiu, G.~Qu, W.~Wang, and M.~Song, ``zkcross: {A} novel architecture for cross-chain privacy-preserving auditing,'' in \emph{33rd {USENIX} Security Symposium, {USENIX} Security 2024, Philadelphia, PA, USA, August 14-16, 2024}, D.~Balzarotti and W.~Xu, Eds.\hskip 1em plus 0.5em minus 0.4em\relax {USENIX} Association, 2024.

\bibitem{Privacy-Preserving_Cryptocurrency_With_Threshold_Authentication_and_Regulation}
Z.~Zhang, C.~Xu, and Y.~Han, ``Privacy-preserving cryptocurrency with threshold authentication and regulation,'' \emph{{IEEE} Trans. Inf. Forensics Secur.}, vol.~19, pp. 6620--6635, 2024.

\bibitem{PGC}
Y.~Chen, X.~Ma, C.~Tang, and M.~H. Au, ``Pgc: Decentralized confidential payment system with auditability,'' in \emph{Computer Security--ESORICS 2020: 25th European Symposium on Research in Computer Security, ESORICS 2020, Guildford, UK, September 14--18, 2020, Proceedings, Part I 25}.\hskip 1em plus 0.5em minus 0.4em\relax Springer, 2020, pp. 591--610.

\bibitem{New_constructions_on_broadcast_encryption_key_pre-distribution_schemes}
S.-H. Huang and D.-Z. Du, ``New constructions on broadcast encryption key pre-distribution schemes,'' in \emph{Proceedings IEEE 24th Annual Joint Conference of the IEEE Computer and Communications Societies.}, vol.~1, 2005, pp. 515--523 vol. 1.

\bibitem{Combinatorial_Bounds_for_Broadcast_Encryption}
M.~Luby and J.~Staddon, ``Combinatorial bounds for broadcast encryption,'' in \emph{Advances in Cryptology - {EUROCRYPT} '98, International Conference on the Theory and Application of Cryptographic Techniques, Espoo, Finland, May 31 - June 4, 1998, Proceeding}, ser. Lecture Notes in Computer Science, K.~Nyberg, Ed., vol. 1403.\hskip 1em plus 0.5em minus 0.4em\relax Springer, 1998, pp. 512--526.

\bibitem{An_efficient_and_secure_multimessage_and_multireceiver_signcryption_scheme}
I.~Ullah, M.~A. Khan, F.~Khan, M.~A. Jan, R.~Srinivasan, S.~Mastorakis, S.~Hussain, and H.~Khattak, ``An efficient and secure multimessage and multireceiver signcryption scheme for edge-enabled internet of vehicles,'' \emph{IEEE Internet of Things Journal}, vol.~9, no.~4, pp. 2688--2697, 2022.

\bibitem{Efficient_anonymous_multireceiver_certificateless_encryption}
Y.-H. Hung, S.-S. Huang, Y.-M. Tseng, and T.-T. Tsai, ``Efficient anonymous multireceiver certificateless encryption,'' \emph{IEEE Systems Journal}, vol.~11, no.~4, pp. 2602--2613, 2017.

\bibitem{RSA}
R.~L. Rivest, A.~Shamir, and L.~Adleman, ``A method for obtaining digital signatures and public-key cryptosystems,'' \emph{Communications of the ACM}, vol.~21, no.~2, pp. 120--126, 1978.

\bibitem{Paillier}
P.~Paillier, ``Public-key cryptosystems based on composite degree residuosity classes,'' in \emph{International conference on the theory and applications of cryptographic techniques}.\hskip 1em plus 0.5em minus 0.4em\relax Springer, 1999, pp. 223--238.

\bibitem{ElGamal}
T.~E. Gamal, ``A public key cryptosystem and a signature scheme based on discrete logarithms,'' in \emph{Advances in Cryptology, Proceedings of {CRYPTO} '84, Santa Barbara, California, USA, August 19-22, 1984, Proceedings}, G.~R. Blakley and D.~Chaum, Eds.

\bibitem{Short_Group_Signatures}
D.~Boneh, X.~Boyen, and H.~Shacham, ``Short group signatures,'' in \emph{Advances in Cryptology - {CRYPTO} 2004, 24th Annual International CryptologyConference, Santa Barbara, California, USA, August 15-19, 2004, Proceedings}, M.~K. Franklin, Ed.

\bibitem{zkLedger}
N.~Narula, W.~Vasquez, and M.~Virza, ``$\{$zkLedger$\}$:$\{$Privacy-Preserving$\}$ auditing for distributed ledgers,'' in \emph{15th USENIX symposium on networked systems design and implementation (NSDI 18)}, 2018, pp. 65--80.

\bibitem{MiniLedger}
P.~Chatzigiannis and F.~Baldimtsi, ``Miniledger: Compact-sized anonymous and auditable distributed payments,'' in \emph{European Symposium on Research in Computer Security}.\hskip 1em plus 0.5em minus 0.4em\relax Springer, 2021, pp. 407--429.

\bibitem{Signatures_on_randomizable_ciphertexts}
O.~Blazy, G.~Fuchsbauer, D.~Pointcheval, and D.~Vergnaud, ``Signatures on randomizable ciphertexts,'' in \emph{Public Key Cryptography - {PKC} 2011 - 14th International Conference on Practice and Theory in Public Key Cryptography, Taormina, Italy, March 6-9, 2011. Proceedings}, ser. Lecture Notes in Computer Science.\hskip 1em plus 0.5em minus 0.4em\relax Springer, 2011, pp. 403--422.

\bibitem{Efficient_signatures_on_randomizable_ciphertexts}
B.~Bauer and G.~Fuchsbauer, ``Efficient signatures on randomizable ciphertexts,'' in \emph{Security and Cryptography for Networks: 12th International Conference, SCN 2020, Amalfi, Italy, September 14--16, 2020, Proceedings 12}.\hskip 1em plus 0.5em minus 0.4em\relax Springer, 2020, pp. 359--381.

\bibitem{BlindHub}
X.~Qin, S.~Pan, A.~Mirzaei, Z.~Sui, O.~Ersoy, A.~Sakzad, M.~F. Esgin, J.~K. Liu, J.~Yu, and T.~H. Yuen, ``Blindhub: Bitcoin-compatible privacy-preserving payment channel hubs supporting variable amounts,'' in \emph{44th {IEEE} Symposium on Security and Privacy, {SP} 2023, San Francisco, CA, USA, May 21-25, 2023}.\hskip 1em plus 0.5em minus 0.4em\relax {IEEE}, 2023, pp. 2462--2480.

\bibitem{Bulletproofs++}
L.~Eagen, S.~Kanjalkar, T.~Ruffing, and J.~Nick, ``Bulletproofs++: next generation confidential transactions via reciprocal set membership arguments,'' in \emph{Annual International Conference on the Theory and Applications of Cryptographic Techniques}.\hskip 1em plus 0.5em minus 0.4em\relax Springer, 2024, pp. 249--279.

\bibitem{GGM}
V.~Shoup, ``Lower bounds for discrete logarithms and related problems,'' in \emph{Advances in Cryptology—EUROCRYPT’97: International Conference on the Theory and Application of Cryptographic Techniques Konstanz, Germany, May 11--15, 1997 Proceedings 16}.\hskip 1em plus 0.5em minus 0.4em\relax Springer, 1997, pp. 256--266.

\bibitem{Proof_systems_for_general_statements_about_discrete_logarithms}
J.~Camenisch and M.~Stadler, ``Proof systems for general statements about discrete logarithms,'' \emph{D-INFK Technical Report}, vol. 260, 1997.

\bibitem{How_to_prove_yourself}
A.~Fiat and A.~Shamir, ``How to prove yourself: Practical solutions to identification and signature problems,'' in \emph{Conference on the theory and application of cryptographic techniques}.\hskip 1em plus 0.5em minus 0.4em\relax Springer, 1986, pp. 186--194.

\bibitem{Security_Arguments_for_Digital_Signatures_and_Blind_Signatures}
D.~Pointcheval and J.~Stern, ``Security arguments for digital signatures and blind signatures,'' \emph{J. Cryptol.}, vol.~13, no.~3, pp. 361--396, 2000.

\bibitem{mcl}
S.~Mitsunari, ``mcl: a portable and fast pairing-based cryptography library,'' \url{https://github.com/herumi/mcl}.

\bibitem{BLS12_381}
S.~Bowe, ``{BLS12-381}: New zk-snark elliptic curve construction,'' \url{https://electriccoin.co/blog/new-snark-curve/}, 2017.

\end{thebibliography}

\end{document}